%% file: ANA-EXOT-2024-21-PAPER.tex
\pdfoutput=1
\documentclass[PAPER, cernpreprint, UKenglish, keywords=true, texmf, orcidlogo]{atlasdoc}
\usepackage[subcaption]{atlaspackage}
\usepackage{acronym}
\usepackage{makecell}

\makeatletter
\newcommand*\AC@undo@cref@label[1]{%
\@bsphack
\protected@write\@auxout{}{%
\string\AC@undonewlabel{#1@cref}%
}%
\@esphack
}
\ifcsname AC@verridelabel\endcsname
\let\AC@orig@verridelabel\AC@verridelabel
\renewcommand*{\AC@verridelabel}[1]{%
\AC@undo@cref@label{#1}%
\AC@orig@verridelabel{#1}%
}
\else
\let\AC@orig@verridelabel\@verridelabel
\renewcommand*{\@verridelabel}[1]{%
\AC@undo@cref@label{#1}%
\AC@orig@verridelabel{#1}%
}
\fi
\makeatother

\usepackage{atlasbiblatex}

\usepackage{atlasphysics}

\graphicspath{{logos/}{figures/}}

\usepackage{ANA-EXOT-2024-21-PAPER-defs}

\AtlasTitle{Search for massive, long-lived particles in events with displaced vertices and displaced muons in $\mathit{pp}$~collisions at $\sqrt{s}=13.6$\,TeV with the ATLAS experiment}

\AtlasAbstract{%
A search is presented for massive long-lived particles in events featuring at least one displaced vertex and at least one displaced muon, using proton--proton collision data collected by the ATLAS detector at the Large Hadron Collider from 2022 to 2024 at a centre-of-mass energy of 13.6\,TeV.
The data sample corresponds to an integrated luminosity of 164\,fb$^{-1}$.
The analysis targets scenarios in which long-lived particles decay inside the ATLAS inner detector, resulting in a topology of at least one massive, displaced vertex~(DV) with multiple associated tracks, and at least one muon with a large transverse impact parameter relative to the primary interaction point.
The muon is not required to be associated with the DV.
Two signal regions are defined by the transverse distance of the reconstructed DV from the interaction point.
Background contributions are estimated by using fully data-driven techniques.
No significant excess above the expected background is observed.
Upper limits at 95\% confidence level are set on the visible cross-section and on the production cross-sections of several benchmark models of $R$-parity-violating supersymmetry.
}

\AtlasRefCode{EXOT-2024-21}

\PreprintIdNumber{CERN-EP-2026-031}

\AtlasKeywords{Long-lived particles, Displaced vertex, Displaced muon, Supersymmetry, R-parity}

\arXivId{2603.01991}

\HEPDataRecord{ins3124901}

\AtlasJournal{\PLB}
\AtlasJournalRef{\PLB 878 (2026) 140509}
\AtlasDOI{10.1016/j.physletb.2026.140509}

\hypersetup{pdftitle={ATLAS document},pdfauthor={The ATLAS Collaboration}}

\begin{document}

\maketitle

%

\section{Introduction}

The \ac{SM} of particle physics is a remarkably successful framework, yet it leaves several fundamental questions unanswered.
It does not provide an explanation for the nature of dark matter, the observed matter--antimatter asymmetry in the Universe, or the origin of neutrino masses.
These and other shortcomings strongly motivate the existence of \ac{BSM} physics.
The \ac{LHC}~\cite{Evans:2008zzb} at CERN is a powerful tool to probe \ac{BSM} scenarios through high-energy proton--proton ($pp$) collisions, enabling searches for new particles and interactions.
While most \ac{BSM} searches at the \ac{LHC} have targeted promptly decaying particles, many extensions of the \ac{SM} predict \acp{LLP} that travel measurable distances before decaying, giving rise to distinctive signatures such as a \ac{DV}~\cite{Alimena:2019zri}.
This Letter presents a search for \acp{LLP} using $pp$ collision data in events with a \ac{DV} and a displaced muon.
\ac{BSM} models resulting in \acp{DV} include various \ac{SUSY} models addressing the hierarchy problem like
minisplit \ac{SUSY}~\cite{Arvanitaki:2012ps, ArkaniHamed:2012gw},
gauge-mediated \ac{SUSY} breaking~\cite{Giudice:1998bp},
\ac{RPV} \ac{SUSY}~\cite{Farrar:1978xj,Golfand:1971iw,Volkov:1973ix,Wess:1974tw,Wess:1974jb,Ferrara:1974pu,Salam:1974ig,Barbier:2004ez,Allanach_2007},
Stealth \ac{SUSY}~\cite{Josh1,Josh2},
and neutral naturalness~\cite{Chacko:2015fbc,Burdman:2006tz, Cai:2008au, Chacko:2005pe},
as well as hidden valley models~\cite{Strassler:2006im,Strassler:2006ri}
and others addressing dark matter~\cite{Baumgart:2009tn, Kaplan:2009ag, Chan:2011aa, Dienes:2011ja, Dienes:2012yz}
and the matter--antimatter asymmetry of the
Universe~\cite{CuiSundrum:2012,Cui:2013,Cui:2014twa}.
Many such models predict final states with displaced muons, which facilitate efficient triggering and provide powerful background suppression.

The results are presented with a model-independent interpretation on the visible cross-section for events with a \ac{DV} and a displaced muon, and with interpretations in the context of \ac{SUSY} models with \ac{RPV} interactions.
Under the \ac{RPV} assumption, the \ac{LSP} is no longer stable and may decay into \ac{SM} particles via \ac{LNV} or \ac{BNV} interactions.
If the corresponding couplings are small, the decays can be significantly suppressed, leading to displaced signatures.
Higgsinos and top squarks (\stopparticle) are common candidates for an unstable \ac{LSP}, usually produced in pairs at the \ac{LHC}, and can decay into final states containing muons and quarks in various \ac{RPV} scenarios.
If the \ac{LSP} is a higgsino, its \ac{RPV} decay proceeds via a virtual sfermion mediator.
If the sfermion mass is assumed to be much heavier than the higgsino mass, this leads to three-body decays in which the virtual sfermion is far off-shell and effectively decouples from the decay kinematics.
In this regime, the higgsino lifetime depends only on the higgsino mass and the \ac{RPV} coupling strength divided by the square of the mediator sfermion mass.
If the \ac{LSP} is the top squark, it decays directly into \ac{SM} particles via the \ac{RPV} coupling.
This two-body decay results in a lifetime that depends on the top-squark mass and the \ac{RPV} coupling strength.

Both the ATLAS and CMS Collaborations have conducted searches for long-lived higgsinos and top squarks in \ac{RPV} \ac{SUSY}~\cite{SUSY-2011-02,SUSY-2012-14,CMS-B2G-12-024,CMS-EXO-16-003,SUSY-2018-33,CMS-EXO-19-013,CMS-EXO-18-003}.
Using Run-1 data at $\sqrt{s} = \SI{8}{\TeV}$, ATLAS searched for higgsino-like neutralinos ($\neutralinoOne$) produced via heavier squarks and decaying through $\neutralinoOne \to \mu^- u\bar{d}$ (and charge conjugate (c.c.)), excluding neutralino masses under this production assumption between \num{100} and \SI{500}{\GeV} for mean proper lifetimes, $\tau$, between \SI{3}{\ps} and \SI{10}{\ns}~\cite{SUSY-2012-14}.
Using Run-2 data at $\sqrt{s} = \SI{13}{\TeV}$, CMS searched for higgsino-like neutralinos with $\neutralinoOne \to tbs$ (and~c.c.) decays in events with two \acp{DV}, excluding neutralino masses up to \SI{1100}{\GeV} for lifetimes between \SI{2}{\ps} and \SI{230}{\ps}~\cite{CMS-EXO-19-013}.
For $\stoppair$ production, ATLAS placed limits on $\stopparticle \to b\mu$ decays in topologies with a \ac{DV} and a displaced muon track using Run-2 data at $\sqrt{s} = \SI{13}{\TeV}$~\cite{SUSY-2018-33}.
Top-squark masses up to \SI{1.7}{\TeV} were excluded for $\tau = \SI{0.1}{\ns}$, and masses below \SI{1.3}{\TeV} were excluded for lifetimes between \SI{0.01}{\ns} and \SI{30}{\ns}.
Most recently, CMS searched for $\stoppair$ production in \ac{RPV} $\stopparticle \to d\ell$ and $\stopparticle \to b\ell$ decays, where $\ell$ denotes an electron or a muon, using Run-2 data at $\sqrt{s} = \SI{13}{\TeV}$~\cite{CMS-EXO-18-003}.
This excluded top-squark masses between \SI{100}{\GeV} and at least \SI{460}{\GeV} for $\SI{0.3}{\ps} < \tau < \SI{30}{\ns}$, with a maximum exclusion of \SI{1500}{\GeV} at $\tau = \SI{70}{\ps}$.

Data from $pp$ collisions at $\sqrt{s} = \SI{13.6}{\TeV}$ collected by the ATLAS detector between 2022 and 2024 of the \ac{LHC} Run~3 are used.
The selected data sample corresponds to an integrated luminosity of \SI{164}{\ifb}~\cite{ATL-DAPR-PUB-2025-001}.
Several techniques novel to Run~3 are employed, including a dedicated displaced-muon trigger~\cite{TRIG-2022-01} extending sensitivity to muon transverse momenta as low as \SI{20}{\GeV}, thereby improving the reach for lower-mass \acp{LLP},
and specialised tracking and vertexing algorithms~\cite{IDTR-2021-03} for reconstructing displaced decay signatures across a broad lifetime range.
The background contributions are estimated by using fully data-driven techniques, and the results are extracted via a profile likelihood fit in a binned observable distribution that separates signal and background.
The results are expressed as model-independent upper limits on the visible cross-section for processes yielding a \ac{DV} and a displaced muon, and are further interpreted in three representative \ac{RPV} \ac{SUSY} scenarios involving higgsino pair production and $\stoppair$ production.


%

\section{ATLAS detector}

\newcommand{\AtlasCoordFootnote}{%
ATLAS uses a right-handed coordinate system with its origin at the nominal interaction point (IP)
in the centre of the detector and the \(z\)-axis along the beam pipe.
The \(x\)-axis points from the IP to the centre of the LHC ring,
and the \(y\)-axis points upwards.
Polar coordinates \((r,\phi)\) are used in the transverse plane,
\(\phi\) being the azimuthal angle around the \(z\)-axis.
The pseudorapidity is defined in terms of the polar angle \(\theta\) as \(\eta = -\ln \tan(\theta/2)\) and is equal to the rapidity
$ y = \frac{1}{2} \ln \left( \frac{E + p_z}{E - p_z} \right) $ in the relativistic limit.
Angular distance is measured in units of \(\Delta R \equiv \sqrt{(\Delta y)^{2} + (\Delta\phi)^{2}}\).
}

The ATLAS experiment~\cite{PERF-2007-01,GENR-2019-02} at the \ac{LHC} is a multipurpose particle detector
with a forward--backward symmetric cylindrical geometry and a near \(4\pi\) coverage in
solid angle.\footnote{\AtlasCoordFootnote}
It consists of an \ac{ID} surrounded by a thin superconducting solenoid
providing a \SI{2}{\tesla} axial magnetic field, \ac{EM} and hadronic calorimeters, and a \ac{MS}.
The \ac{ID} covers the pseudorapidity range \(|\eta| < 2.5\).
It consists of silicon pixel, \ac{SCT}, and transition radiation tracking (TRT) detectors.
Lead/\ac{LAr} sampling calorimeters provide \ac{EM} energy measurements
with high granularity within the region \(|\eta|< 3.2\).
A steel/scintillator-tile hadronic calorimeter covers the central pseudorapidity range (\(|\eta| < 1.7\)).
The endcap and forward regions are instrumented with \ac{LAr} calorimeters
for \ac{EM} and hadronic energy measurements up to \(|\eta| = 4.9\).
The \ac{MS} surrounds the calorimeters and is based on
three large superconducting air-core toroidal magnets with eight coils each.
The field integral of the toroids ranges between \num{2.0} and \SI{6.0}{\tesla\metre}
across most of the detector.
The \ac{MS} includes three precision layers, each consisting of layers of monitored drift tubes, covering the region $|\eta| < 2.7$, except in the innermost layer of the endcap region, where layers of small thin-gap and Micromegas chambers provide both precision tracking in the region $1.3<|\eta|<1.7$.
The muon trigger system, covering $|\eta|<2.4$, uses various detectors including resistive-plate chambers in the barrel region, thin-gap chambers in the endcaps, and small-strip thin-gap chambers together with Micromegas detectors in the innermost endcap layer.
The luminosity is measured mainly by the LUCID--2 detector~\cite{LUCID2}, which is located close to the beampipe.
A two-level trigger system was used to select events~\cite{TRIG-2022-01}.
The first-level trigger is implemented in hardware and uses a subset of the detector information
to accept events at a rate close to \SI{100}{\kHz}.
This is followed by a software-based \ac{HLT} that
reduces the accepted rate of complete events to \SI{3}{\kHz} on average,
depending on the data-taking conditions.
A software suite~\cite{SOFT-2022-02} is used in data simulation, in the reconstruction
and analysis of real and simulated data, in detector operations, and in the trigger and data acquisition
systems of the experiment.


%

\section{Samples of simulated events}
\label{sec:samples}

\Ac{MC} simulation samples are used to study the event selection acceptance and the reconstruction efficiency for benchmark signals.
The effect of multiple interactions in the same and neighbouring bunch crossings (pile-up) is modelled by overlaying~\cite{SIMU-2020-01} the simulated hard-scattering event with inelastic $pp$ events generated from a mix of \EPOS[2.0.1.4]~\cite{Werner:2005jf} and \PYTHIA[8.308]~\cite{Sjostrand:2014zea,Bierlich:2022pfr}.
The detector response to each simulated event was modelled using \GEANT~\cite{Agostinelli:2002hh}.
Correction factors are applied to account for differences between data and simulation in the pile-up, the trigger, reconstruction, and identification efficiencies of muons~\cite{TRIG-2022-01,MUON-2018-03}, and for the track reconstruction efficiency~\cite{IDTR-2021-03}.

Samples of simulated long-lived higgsino pairs are used as benchmarks to set limits on the visible cross-section for higgsino pair production.
The following production modes are considered:
$\neutralinoOne\neutralinoTwo$,
$\neutralinoOne\chargino$,
$\neutralinoTwo\chargino$,
and
$\chargino\charginoInv$,
where \neutralinoOne is the \ac{LSP}, and \neutralinoTwo and \chargino are the next-to-lightest neutralino and the lightest chargino, respectively.
A nearly mass-degenerate spectrum is assumed, with mass splittings of \SI{1}{\GeV} between \neutralinoOne and \chargino, and \neutralinoOne and \neutralinoTwo~\cite{Ibe_2013,Nagata_2015}.
All other supersymmetric particles are assumed to be heavy and decoupled.

Two \ac{RPV} decay scenarios are considered, where the higgsinos are assumed to decay through interactions that are either \ac{LNV} or \ac{BNV}~\cite{Barbier:2004ez}.
These interactions arise from additional terms in the supersymmetric superpotential of the form $\lambda'_{ijk} L_i Q_j D^c_k$ and $\lambda''_{ijk} U^c_i D^c_j D^c_k$, where $\lambda'_{ijk}$ and $\lambda''_{ijk}$ are the coupling strengths and $i$, $j$, and $k$ are generation indices.
Here, $L$ and $Q$ denote the lepton and quark $SU(2)$ doublet superfields, while $U^c$ and $D^c$ are the up- and down-type quark $SU(2)$ singlet superfields.
Under the decoupling assumption, the masses of the sfermions that mediate the higgsino decays are much larger than the higgsino masses.
Thus, the higgsino decays are modelled as effective three-body decays via a contact interaction, with the higgsino mass and decay width treated as independent simulation parameters, without resolving the underlying dependence on the \ac{RPV} coupling strength and mediator sfermion mass.
The following two higgsino benchmark scenarios were simulated, both assuming that all other \ac{RPV} terms are zero:
\begin{itemize}
\item \ac{LNV}: $\lambda'_{211}$, leading to decays $\neutralinoOne, \neutralinoTwo \to \mu^- u\bar{d}$ (+~c.c.), and $\chargino \to \mu^{\pm} d\bar{d}$.
\item \ac{BNV}: $\lambda''_{323}$, leading to decays $\neutralinoOne, \neutralinoTwo \to tsb$ (+~c.c.), and $\charginoplus \to \bar{b}\bar{s}\bar{b}$, $\charginominus \to bsb$.
Sensitivity for this scenario depends on neutralino production and the subsequent semileptonic decay of the resulting top quark, leading to a lower muon multiplicity and reduced signal acceptance with respect to the \ac{LNV} benchmark.
\end{itemize}%
\begin{figure}
\centering
\subcaptionbox{\label{fig:feynman_diagrams_lqd}}{\includegraphics[width=0.33\textwidth,clip=true,trim=35pt 5pt 5pt 5pt]{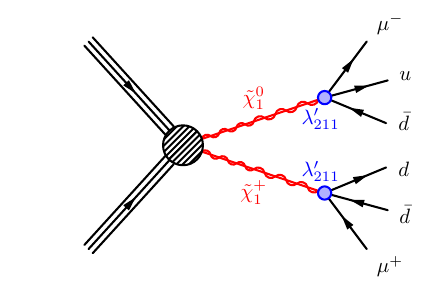}}%
\hfill
\subcaptionbox{\label{fig:feynman_diagrams_udd}}{\includegraphics[width=0.33\textwidth,clip=true,trim=35pt 5pt 5pt 5pt]{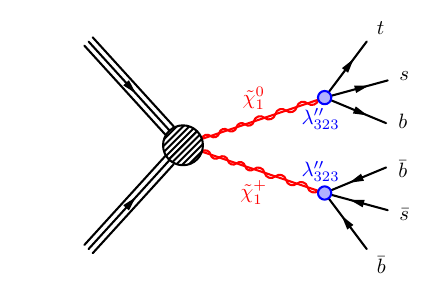}}%
\hfill
\subcaptionbox{\label{fig:feynman_diagrams_top}}{\includegraphics[width=0.33\textwidth,clip=true,trim=35pt 5pt 5pt 5pt]{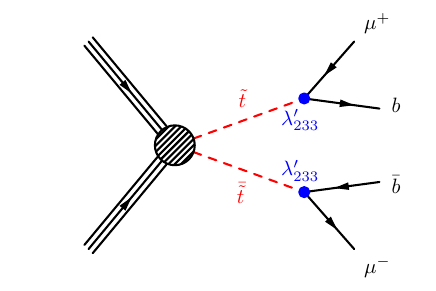}}%
\caption{%
Illustrative diagrams for the three considered \acf{RPV} \ac{SUSY} simplified benchmark models, each assuming a single non-zero RPV term:
\subref{fig:feynman_diagrams_lqd}~higgsino pair production with $\lambda'_{211}$ decays,
\subref{fig:feynman_diagrams_udd}~higgsino production with $\lambda''_{323}$ decays, and
\subref{fig:feynman_diagrams_top}~top-squark--anti-squark pair production with $\lambda'_{233}$ decays.
Equivalent diagrams with charge-conjugate processes are also considered.
}
\end{figure}
The decay modes are illustrated in \cref{fig:feynman_diagrams_lqd,fig:feynman_diagrams_udd}.
The hard interaction was generated with up to two additional partons with the \MGNLO[3.5.7]~\cite{Alwall:2014hca} generator at \ac{LO} in perturbative \ac{QCD}, using the \NNPDF[3.0nlo] \ac{PDF} set~\cite{Ball:2014uwa}.
The parton shower and hadronisation were simulated with \Pythia[8.308] using the A14 set of tuned parameters~\cite{ATL-PHYS-PUB-2014-021} and the \NNPDF[2.3lo] \ac{PDF} set~\cite{Ball:2012cx}.
The decays of bottom and charm hadrons were simulated by \EVTGEN[1.6.0]~\cite{Lange:2001uf}.
The \ac{LNV} (\ac{BNV}) decay mode was generated for higgsino lifetimes between \SI{3}{\pico\second} and \SI{100}{\nano\second} for a range of mass values of \num{100}--\SI{1900}{\GeV} (\num{200}--\SI{1500}{\GeV}).
The generated sets of events were scaled to cross-sections at approximate \ac{NNLO} accuracy in \ac{QCD} with threshold resummation of the \acp{NNLL}~\cite{Fiaschi:2020udf}, as computed with the \AtlasMC{Resummino} package~\cite{Fuks:2013vua,Fiaschi:2023tkq}.
The resulting production cross-sections decrease rapidly with increasing higgsino mass, from $\SI{19.5}{\pb}$ at $m = \SI{100}{\GeV}$ to $\mathcal{O}(\SI{e-2}{\fb})$ at $m = \SI{1900}{\GeV}$, with scale uncertainties of a few percent and \ac{PDF} uncertainties increasing from a few percent at low masses to $\mathcal{O}(50\text{--}100\%)$ at the largest masses considered.
Interactions of charginos with the detector material were found to have a negligible impact on the signal yields and were not simulated in the \ac{MC} samples.

Using the same \MGNLOPY generator, \stoppair production samples were generated at \ac{LO} \ac{QCD} with up to two additional partons.
The other top-squark mass eigenstate and all other supersymmetric particles are assumed to be heavy and decoupled.
The squarks are assumed to decay exclusively via \ac{LNV} couplings to a muon and a $b$-quark, $\stopparticle \to \mu b$, given by a non-zero $\lambda'_{233}$ coupling term.
The diagram for this process is illustrated in \cref{fig:feynman_diagrams_top}.
Different samples were generated with $\SI{3}{\pico\second} < \tau < \SI{100}{\nano\second}$ for a range of mass values between \SI{700}{\GeV} and \SI{2100}{\GeV}.
The generated sets of events were scaled to cross-sections computed at approximate \ac{NNLO} accuracy in \ac{QCD}.
The production cross-section decreases rapidly with increasing top-squark mass, from about $\SI{0.1}{\pb}$ at $m_{\stopparticle} = \SI{700}{\GeV}$ to about $\SI{e-2}{\fb}$ at $m_{\stopparticle} = \SI{2100}{\GeV}$, with associated theoretical uncertainties increasing from roughly 6\% to about 36\%.
Since the top squarks carry colour charge and all simulated proper lifetimes are far longer than the hadronisation timescale in \ac{QCD}, they were hadronised in \Pythia[] to form colour-singlet states with \ac{SM} quarks, called \emph{R-hadrons}.
The top squarks primarily form meson-like states ($\stopparticle\bar{q}$), but approximately 10\% of them form baryon-like states ($\stopparticle qq$).
About half of the R-hadrons carry non-zero electric charge.
Interactions of the R-hadrons with the detector material were not simulated in the \ac{MC} samples.


%

\section{Reconstruction}

Charged-particle trajectories (tracks) are reconstructed in two passes using \emph{hits} (individual detector measurement points) found in the \ac{ID}.
The primary pass reconstructs tracks with transverse impact parameter (\dzero) relative to the \ac{IP} of $|\dzero| < \SI{5}{\mm}$ and a minimum transverse momentum (\pt) of \SI{500}{\MeV}~\cite{IDTR-2022-04}.
A large-impact-parameter pass uses left-over hits from the primary pass to increase the acceptance up to $|\dzero| < \SI{300}{\mm}$ and a minimum \pt of \SI{1}{\GeV}~\cite{IDTR-2021-03}.
All tracks are passed through an ambiguity resolution step to resolve overlaps and remove fake tracks
before being re-fit using a global $\chi^2$ method~\cite{IDTR-2021-03}. The tracks from the silicon pixel and \ac{SCT} sub-detectors may then be extended into the TRT~\cite{IDTR-2022-04}.
The $pp$ vertex with the highest sum of squared transverse momenta of associated tracks is taken as the \ac{PV}.

Muons are reconstructed using the standard ATLAS reconstruction suite~\cite{IDTR-2022-04} by combining both primary and large-impact-parameter \ac{ID} tracks with track segments reconstructed in the \ac{MS} within $|\eta|<2.5$; the performance of this combined approach under the 2022 and 2023 Run~3 data-taking conditions was studied in a previous ATLAS analysis~\cite{SUSY-2022-11}.
If an \ac{MS} track segment is matched to both a primary and large-impact-parameter track, the combined track with the best combined fit quality for the given \ac{MS} track is retained. If they have the same quality, the muon with the smallest $\Delta\eta$ between the MS and ID tracks is kept~\cite{IDTR-2022-04}.
Muons must satisfy a modified version of the \textit{Medium} identification criteria as defined in Ref.~\cite{MUON-2018-03}, where the requirements on the number of pixel and SCT hits are omitted if reconstructed from large-impact-parameter tracks.
Only muons with $\pt > \SI{25}{\GeV}$, $\SI{2}{\mm} < |\dzero| < \SI{300}{\mm}$,
a transverse-displacement significance $|\dzero|/\sigma_{\dzero} > 150$, and a longitudinal displacement relative to the \ac{PV} of $|z_0| < \SI{500}{\mm}$ are considered.

To reconstruct the vertex of the \ac{LLP} decay, primary and large-impact-parameter tracks are passed to a modified version of the \ac{DV} reconstruction algorithm described in Ref.~\cite{ATL-PHYS-PUB-2019-013}.
The vertex reconstruction first forms two-track vertices from a set of \emph{seed tracks}.
These seed tracks are selected following the criteria in Ref.~\cite{ATL-PHYS-PUB-2019-013}, but with the requirement for at least one track in the seed pair to have $|\dzero| > \SI{2}{\mm}$ relaxed to $|\dzero| > \SI{1}{\mm}$, improving efficiency for \ac{LLP} decays whose decay products -- particularly in models with short lifetimes -- tend to point back towards the \ac{PV} and have small $|\dzero|$.
Two-track vertices are combined to form multi-track vertices and tracks not initially preselected for vertex seeding are attached to compatible vertices.

Candidate \acp{DV} are required to be within the fiducial volume of the \ac{ID}, with a transverse distance from the beamline of $R_{xy} < \SI{300}{\mm}$ and a longitudinal position of $|z| < \SI{300}{\mm}$,
but with a minimum transverse distance from the \ac{PV} of $\dTPVDV > \SI{1}{\mm}$.
They are further required to fulfil a vertex quality requirement of $\chi^2_{\text{red}} < 5$, where \enquote{red} in the subscript indicates division by the number of degrees of freedom of the vertex fit, to suppress vertices formed from random track crossings.
Additional background rejection is achieved using two angular requirements that exploit the expected topology of displaced decays.
These requirements suppress backgrounds in which tracks are misassociated with the \ac{DV} or arise from random crossings or back-to-back configurations.
First, a minimum azimuthal angle between the \ac{DV} displacement vector and any of its associated tracks is required, $\cosminphi > -0.995$, rejecting vertices containing tracks that point nearly opposite to the \ac{DV} displacement direction.
Second, among all tracks forming an acute angle with the \ac{DV} displacement direction, at least one must have a sufficiently large angle, $\cosminphiacute < 0.97$.
This requirement suppresses backgrounds where tracks are nearly collinear with the displacement direction, while retaining signal-like decays with a broader angular spread.


%

\section{Event selection}

\begin{figure}[tp]
\centering
\subcaptionbox{\label{fig:trig_eff_muon_pT}}{\includegraphics[width=0.49\textwidth,clip=true,trim=10pt 15pt 10pt 15pt]{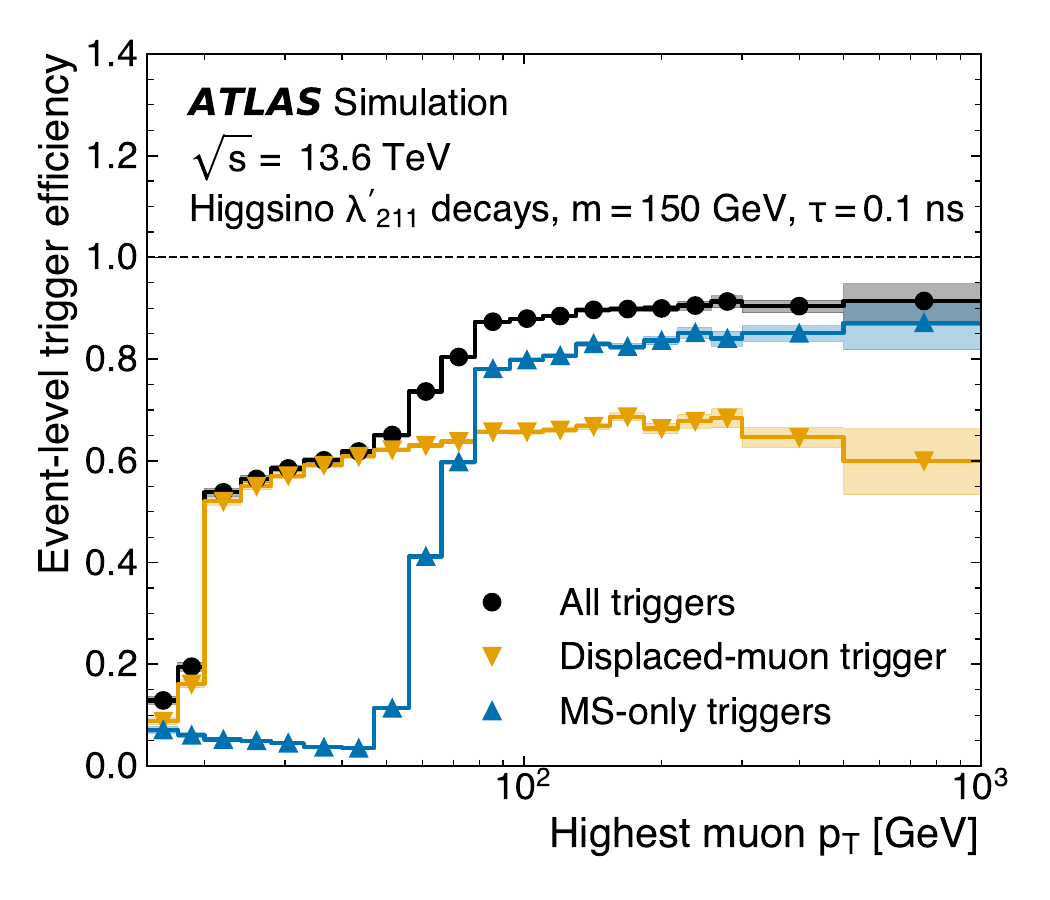}}%
\hfill%
\subcaptionbox{\label{fig:trig_eff_muon_d0}}{\includegraphics[width=0.49\textwidth,clip=true,trim=10pt 15pt 10pt 15pt]{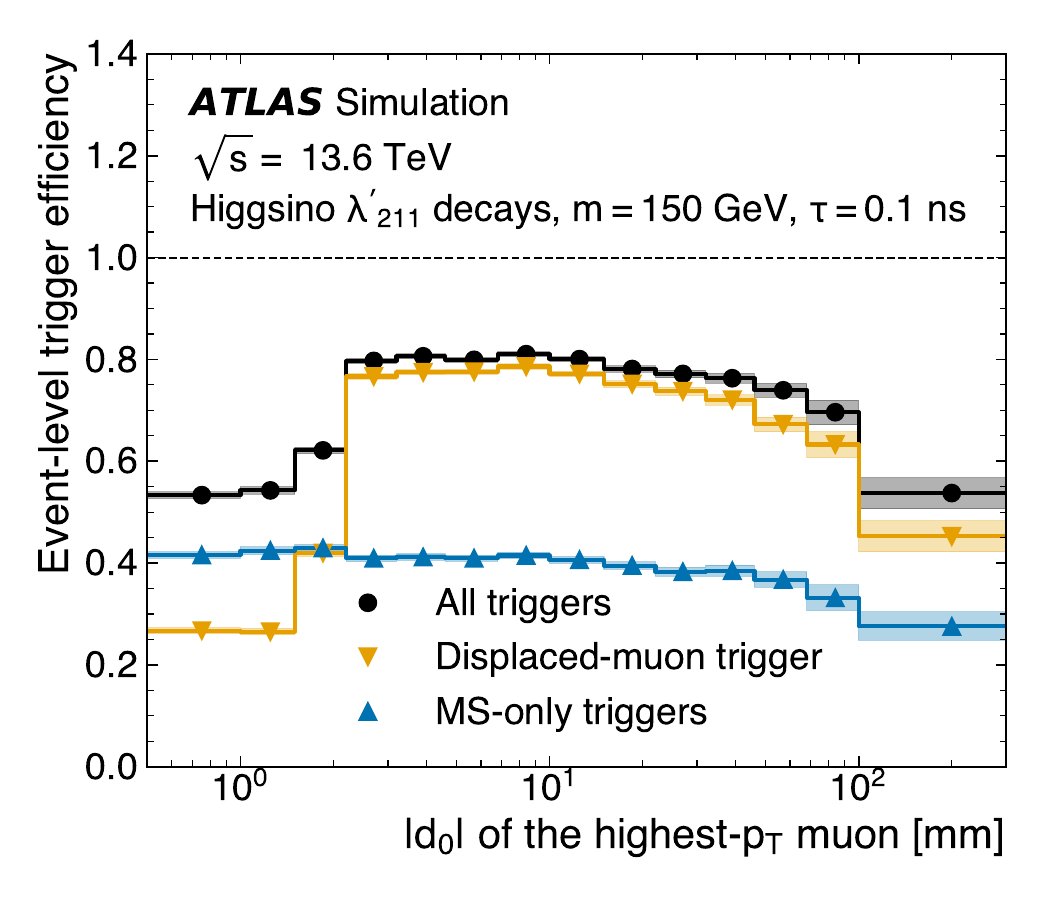}}
\caption{%
Event-level trigger efficiency for the triggers used in this analysis.
The plots show separate curves for the \ac{MS}-only triggers and the displaced-muon trigger, which exploits the dedicated large-impact-parameter track reconstruction introduced in the HLT in Run~3, and their combination.
The efficiencies are evaluated as a function of \subref{fig:trig_eff_muon_pT}~the highest muon $\pt$ in the event, and \subref{fig:trig_eff_muon_d0}~the transverse impact parameter, $|d_0|$, of that muon.
The benchmark sample used here consists of higgsino pair-production events with \ac{LNV} $\lambda'_{211}$ decays ($m=\SI{150}{\GeV}$, $\tau=\SI{0.1}{\ns}$).
For the \ac{MS}-only triggers, the trigger plateau criteria are applied to the highest-\pt muon in the event (\SI{63}{\GeV} for the $\pt > \SI{60}{\GeV}$, $|\eta| < 1.05$ trigger and \SI{84}{\GeV} for the $\pt > \SI{80}{\GeV}$, $|\eta| < 2.4$ trigger with additional requirement of three hits in precision tracking layers for $|\eta| > 1.05$).
The muon \pt plot shows the combined trigger efficiency turn-on behaviour.
The $|d_0|$ distribution reflects the $|d_0| > \SI{2}{\mm}$ requirement of the displaced-muon trigger, with reduced efficiency at large $|d_0|$ due to the limited tracking efficiency for highly displaced muons.
}
\label{fig:trig_eff}
\end{figure}

Events are selected using two \ac{MS}-only triggers and a novel displaced-muon trigger introduced in Run~3~\cite{TRIG-2022-01}.
The \ac{MS}-only triggers accept events based on muon tracks reconstructed in the \ac{MS} without requiring a matching \ac{ID} track.
The first MS-only trigger selects muon track segments with transverse momentum of $\pt > \SI{60}{\GeV}$ for $|\eta| < 1.05$, while the second requires $\pt > \SI{80}{\GeV}$ over the full pseudorapidity range in the muon trigger system of $|\eta| < 2.4$, with an additional requirement of three hits in precision tracking layers for $|\eta| > 1.05$.
The displaced-muon trigger selects events containing at least one muon candidate reconstructed in the HLT with $p_\mathrm{T} > \SI{20}{\GeV}$, $|\eta| < 2.4$,  and $|\dzero| > \SI{2}{\mm}$.
It exploits the dedicated large-impact-parameter track reconstruction introduced in the HLT in Run~3, which is designed to suppress incorrectly reconstructed displaced tracks and enables lower \pt thresholds than in Run~2.
\Cref{fig:trig_eff} shows the trigger efficiency as a function of the highest muon \pt in the event and of its transverse impact parameter, $|d_0|$,
evaluated using simulated higgsino pair-production events with \ac{LNV} $\lambda'_{211}$ decays, $m = \SI{150}{\GeV}$ and $\tau = \SI{0.1}{\ns}$.
The combination of these triggers achieves high efficiency across a broad range of muon \pt and $|d_0|$ values.

Events selected in the analysis must contain at least one muon candidate that either is spatially matched to the HLT muon that fired the displaced-muon trigger, or has \pt at least 5\% above the relevant MS-only trigger threshold (\SI{63}{\GeV} and $\SI{84}{\GeV}$) to operate on the trigger-efficiency plateau~\cite{SUSY-2022-11}.

If present, \acp{DV} are categorised as \enquote{far} or \enquote{near} by their transverse distance from the \ac{PV}, \dTPVDV\@.
Far \acp{DV} have $\dTPVDV \geq \SI{4}{\mm}$ and near \acp{DV} have $\SI{1}{\mm} < \dTPVDV < \SI{4}{\mm}$.
A~\ac{DV} is considered \emph{signal-like} if it has at least four associated tracks ($\ntrack \geq 4$) and if the invariant mass of all \ac{DV}-associated tracks fulfils $\mDV \geq \SI{20}{\GeV}$ for far \acp{DV} and $\mDV \geq \SI{40}{\GeV}$ for near \acp{DV}.
Signal-like \acp{DV} are also required to be outside the material layers defined by a map described in Ref.~\cite{SUSY-2018-13} to veto potential hadronic interactions between particles and the detector material.
This material-map veto removes 42\% of the fiducial volume.
The presence of far or near signal-like \acp{DV} is used to categorise events as detailed in \cref{sec:background}.

Three primary muon background sources are considered: muons from heavy-flavour decays, cosmic-ray muons, and algorithmic fakes.
Heavy-flavour backgrounds are suppressed by requiring candidates to satisfy the \emph{Tight\_VarRad} isolation criteria described in Ref.~\cite{MUON-2018-03}, hereafter referred to as the \emph{heavy-flavour veto}.
This requirement eliminates approximately 98.9\% of heavy-flavour muons and provides an optimal balance between background rejection and signal efficiency.
Cosmic-ray rejection is performed by requiring $|\eta_\mu + \eta_{\mathrm{MS}}| > 0.014$ and $|\phi_\mu - \phi_{\mathrm{MS}} | - \pi > 0.25$, where the $\mu$ (MS) subscripts refer to the reconstructed muon (MS track segment).
This rejects muon candidates with an \ac{MS} track segment on the opposite side of the detector.
A geometric correction is applied so that the muon direction is expressed in the detector reference frame for comparison with the MS track segment.
The $\eta$ and $\phi$ requirements are optimised using a dedicated cosmic-ray data sample recorded in 2023.
Additionally, a muon-material-map veto described in Ref.~\cite{SUSY-2022-11} is applied to account for detector inefficiencies in regions where \ac{MS} track segment reconstruction is not possible.
The combined \emph{cosmic-ray veto} achieves a rejection of approximately 97.5\% for cosmic-ray muons.
Algorithmic fakes -- reconstructed muon candidates that do not originate from real muons -- are suppressed by requiring a high-quality match between the \ac{ID} and \ac{MS} tracks:
the combined fit must satisfy $\chi^2_{\text{red}} < 3$ and have at least three precision layer hits, achieving rejection rates of approximately 97.4\% and 99.7\% in the barrel and endcap regions, respectively.
The combination of these matching requirements is referred to as the \emph{algorithmic-fake veto}.

Events are considered to contain a signal-like muon if the muon with the highest \pt in the event passes the heavy-flavour, cosmic-ray, and algorithmic-fake vetoes.
No association between the signal-like muon and the \ac{DV} is required to preserve model independence.
Two signal regions, denoted \SRfar and \SRnear, are defined and correspond to events with a signal-like muon and at least one far or near signal-like \ac{DV}, respectively.
Only a single \ac{DV} is required to retain sensitivity for long lifetimes, including cases where one decay occurs outside the \ac{ID}, while preserving a model-independent interpretation.
Events in which either the muon or the vertex is not signal-like are placed in validation regions for the background estimate.


%

\section{Background estimate}
\label{sec:background}

Background contributions arise from events with displaced muons from heavy-flavour decays, cosmic rays, or algorithmic fakes, and from \acp{DV} produced by hadronic interactions or random crossings of tracks in the detector.
These backgrounds are efficiently suppressed by the muon-veto requirements and by the requirements on \mDV and \ntrack in the signal regions.
To predict the event yields from these sources in the signal regions, a fully data-driven \acl{TF} method is employed.
In this approach, separate transfer factors are derived for each muon background source~$i$.
The muon-veto requirements and the \ac{DV} selection criteria are used to define exclusive regions in data:
region~$\mathrm{A}$, where both the muon-veto and \ac{DV} selection are applied (the signal region);
regions~$\mathrm{B}_i$ and~$\mathrm{C}$, where exactly one criterion is inverted;
and region~$\mathrm{D}_i$, where both criteria are inverted.
Here, $i$~denotes the specific muon background type.
This corresponds to separate ABCD constructions for each background source, which share the same regions~$\mathrm{A}$ and~$\mathrm{C}$, but have distinct regions~$\mathrm{B}_i$ and~$\mathrm{D}_i$.
The background estimate relies on the fact that the variables used to reject displaced muons from background sources are uncorrelated with the variables used to reject \acp{DV} from background.
The event yields in region~$\mathrm{A}$ for source~$i$ can then be estimated from the observed yields in region~$\mathrm{B}_i$ and the ratio of yields in regions~$\mathrm{C}$ and~$\mathrm{D}_i$, where region~$\mathrm{C}$ is common to all background sources, via
$N_{\mathrm{A},i} = N_{\mathrm{B}_i} \cdot N_{\mathrm{C}} / N_{\mathrm{D}_i}$.
The ratio $f_i = N_{\mathrm{C}} / N_{\mathrm{D}_i}$ is referred to as the \ac{TF} for background source~$i$.
For background events, the probabilities to satisfy the \ac{DV} and muon selection criteria were validated to be independent in data by confirming that the shapes of the muon observables are unchanged under varying \ac{DV} selections.
Any residual correlations are measured and treated as a systematic uncertainty in the background prediction, as described in \cref{sec:tf-uncertainties}.

Potential signal contamination in regions~$\mathrm{B}_i$, $\mathrm{C}$, and~$\mathrm{D}_i$ is evaluated using simulated signal samples.
The expected signal yields are normalised to values consistent with existing exclusion limits~\cite{SUSY-2022-11}.
Under this assumption, the fraction of signal events entering these regions is found to be negligible compared to the background yields.

\begin{figure}[tp]
\centering
\includegraphics[width=\textwidth]{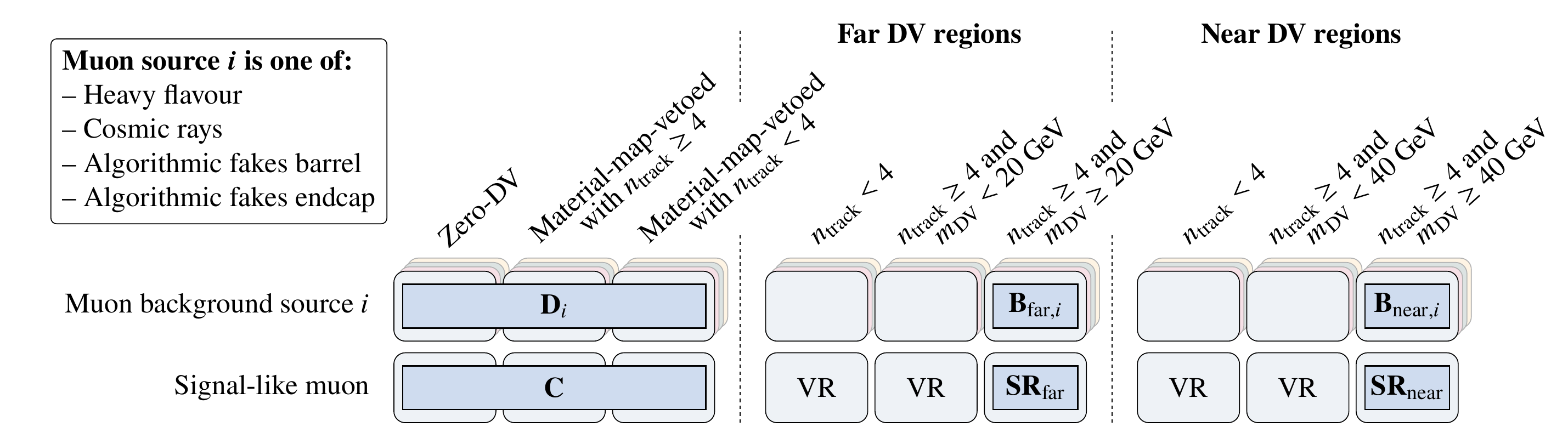}
\caption{%
Categorisation of events based on the presence and quality of \acfp{DV} (horizontal axis) and the background source~$i$ of the highest-\pt muon (vertical axis).
The schematic is compressed vertically for simplicity.
If a muon fails the veto for a specific muon background source, it is assigned that background category;
if it passes all muon requirements, it is considered signal-like.
The signal regions, \SRfar and \SRnear, require at least one signal-like muon and at least one signal-like far or near \ac{DV}, respectively.
Additional validation regions are defined by dropping or inverting the \ntrack and \mDV requirements for signal-like \acp{DV}.
The regions~$\mathrm{C}$ and~$\mathrm{D}_i$ consist of the zero-\ac{DV} and material-map-vetoed \ac{DV} classes; the \acfp{TF} are derived from subsets of these regions by applying additional muon selections to enhance the purity for each source.
The background predictions for \SRfar and \SRnear are then obtained by applying the \acp{TF} to the $\mathrm{B}_{\text{far},i}$ and $\mathrm{B}_{\text{near},i}$ regions, respectively.
}
\label{fig:abcd-schematic}
\end{figure}

\acp{TF} are derived separately for each muon background source~$i$, with algorithmic fakes further split into barrel and endcap categories with a transition at $|\eta| = 1.2$ to account for different fake rates in the two \ac{MS} regions, yielding four muon background sources in total.
Events are categorised based on the highest-\pt muon: if this muon fails exactly one of the three vetoes (heavy-flavour, cosmic-ray, or algorithmic-fake), it is considered to be of that background type and the event enters region~$\mathrm{B}_i$ or~$\mathrm{D}_i$; for algorithmic fakes, the $|\eta|$ of the muon is used to determine the barrel/endcap category.
Regions~$\mathrm{A}$ or~$\mathrm{C}$ consist of events where the highest-\pt muon passes all vetoes.
Events where the highest-\pt muon fails two or more vetoes are discarded.
A schematic overview of the constructed regions is shown in \cref{fig:abcd-schematic}.

For the \ac{DV} selection criterion, events are assigned to categories following a fixed sequence: events with no reconstructed \ac{DV} enter the zero-\ac{DV} category; if a \ac{DV} is present, those failing to meet the \mDV requirement are considered first, followed by those obtained by dropping the \mDV requirement and selecting $\ntrack < 4$, with far \acp{DV} given precedence over near ones at both steps; events with \acp{DV} rejected by the material-layer map, referred to as \emph{material-map-vetoed} \acp{DV}, are further divided into those with $\ntrack \geq 4$ and $\ntrack < 4$.

While material-map-vetoed \acp{DV} originate from hadronic interactions and differ from signal-like \acp{DV}, the efficiency of the muon vetoes in these events is expected to be uncorrelated with whether a \ac{DV} occurs in the detector material or in the fiducial volume.
Consequently, events with material-map-vetoed \acp{DV} and those with no \acp{DV} are used to populate the regions~$\mathrm{C}$ and~$\mathrm{D}_i$, providing high-statistics samples.
These regions contain an admixture of the muon background sources, thus, the \acp{TF} are derived from subsets of these regions, with additional muon requirements to enhance the purity for specific backgrounds, as detailed in \cref{sec:tf-derivation}.
The remaining \ac{DV} categories are used to define validation regions and the signal regions \SRfar and \SRnear.
The total predicted background yield in \SRfar (\SRnear) is obtained as
$N_{\mathrm{A}}^{\text{pred}} = \sum_{i=1}^{4} f_i \, N_{\mathrm{B}_{\text{far},i}}$
($N_{\mathrm{B}_{\text{near},i}}$),
where the sum runs over the four muon background sources~$i$.
To obtain a prediction in \SRfar and \SRnear for a binned observable, the \acp{TF} are multiplied by the corresponding binned yields in $\mathrm{B}_{\text{far},i}$ and $\mathrm{B}_{\text{near},i}$, respectively.

\subsection{Transfer-factor derivation}
\label{sec:tf-derivation}

Zero-\ac{DV} events and those with material-map-vetoed \acp{DV} are highly enriched in background muons, but the \ac{TF} derivation requires sets of events dominated by a single source of muon background.
Thus, additional muon selection criteria are applied to the highest-\pt muon in the event when defining regions~$\mathrm{C}$ and~$\mathrm{D}_i$ for the \ac{TF} calculation:
\begin{enumerate}
\item Muons from heavy-flavour decays dominate at small displacement, so a requirement of \mbox{$|d_0(\mu)| < \SI{3}{\mm}$} is imposed when deriving the heavy-flavour \ac{TF};
for the other three background \acp{TF}, \mbox{$|d_0(\mu)| > \SI{15}{\mm}$} is required to remove heavy-flavour contributions.
These requirements are applied in addition to the overall muon selection criterion of $\SI{2}{\mm} < |d_0(\mu)| < \SI{300}{\mm}$.
\item Cosmic-ray muons predominantly occur in the central detector and enter from one side in azimuth; therefore, $\abseta < 1.05$ and $\phi < 0$ are required to isolate a pure sample for the cosmic-ray \ac{TF} derivation.
\item Muons from algorithmic fakes are treated separately in the barrel ($|\eta| \leq 1.2$) and endcap ($|\eta| > 1.2$) regions; to avoid cosmic-ray contamination, an additional $\phi > 0$ requirement is applied for the barrel.
\end{enumerate}

The derived \acp{TF} are of the order of $\mathcal{O}(10^{-2})$, reflecting the strong suppression of the background muon sources by the muon-veto requirements.
Statistical uncertainties from the finite event yields in the \ac{TF} derivation regions range between 2\% and 6\%.

\subsection{Systematic uncertainties in the transfer factors}
\label{sec:tf-uncertainties}

Multiple uncertainties are assigned to the \acp{TF} to account for potential residual correlations in the probabilities for background events to satisfy the muon and \ac{DV} selection criteria, for residual correlations between the \acp{TF} and the muon properties, and for possible contamination of the \ac{TF} derivation regions by other muon background sources.

Potential residual correlations between the \ac{DV} and muon selection criteria are assessed by recomputing the \acp{TF} individually for zero-\ac{DV} events and for events with \acp{DV} rejected by the material-map veto.
As the latter are enriched in background \acp{DV} from hadronic interactions with detector material, comparing these \acp{TF} provides a test of whether the probabilities to satisfy the muon and \ac{DV} selection criteria are independent for background events.
The largest relative difference between the nominal and recomputed \acp{TF} is assigned as a systematic uncertainty for each muon background class.
These uncertainties range between 8\% and 35\% and are largest for the heavy-flavour and the algorithmic fake barrel \acp{TF}.

To account for residual dependencies of the \ac{TF} values on the muon $|d_0|$, the chosen $|d_0|$ ranges are split into two subregions at midpoint values that balance the statistical uncertainties on either side.
The \acp{TF} are recomputed independently in these subregions and compared with the nominal values.
This comparison provides a check for unmodelled $|d_0|$ dependencies given the available data, without attempting to model a detailed functional dependence.
The largest relative differences between the nominal and recomputed \acp{TF} are assigned as a systematic uncertainty for each background class.
The computed midpoints are $|d_0| = \SI{2.3}{\mm}$ for muons from heavy-flavour decays, $|d_0| = \SI{85}{\mm}$ for cosmic-ray muons, $|d_0| = \SI{130}{\mm}$ for algorithmic fakes in the barrel, and $|d_0| = \SI{102.5}{\mm}$ for algorithmic fakes in the endcap.
The $d_0$ uncertainties range between 3\% for cosmic-ray muons and 21\% for algorithmic fakes in the endcap.

Additional sources of systematic uncertainty are considered for the cosmic-ray, algorithmic-fake barrel and algorithmic-fake endcap \acp{TF} to account for residual dependencies of the \ac{TF} values on the muon $\eta$ and $\phi$ coordinates, and for possible contamination of the \ac{TF} derivation regions by other muon background sources.
For all three \acp{TF}, the $\phi$ regions are split in half to compute separate \acp{TF}, and the largest relative differences between the original and recomputed \acp{TF} are taken as systematic uncertainties.
For the cosmic-ray and algorithmic-fake barrel \acp{TF}, the midpoint is picked so that the statistical uncertainty is minimal on either side, while for the algorithmic-fake endcap \ac{TF}, a split at $\phi = 0$ is used to account for possible residual cosmic-ray contamination in the negative-$\phi$ region.
The cosmic-ray and algorithmic-fake barrel $\phi$ uncertainties amount to approximately 4\%, while the algorithmic-fake endcap $\phi$ uncertainty is as high as 49\%.
For the cosmic-ray \ac{TF}, an additional midpoint split in $\abseta$ that minimises the statistical uncertainties on either side is performed, yielding an additional \abseta uncertainty of 42\%.

\Cref{tab:result_yields} lists the predicted event yields for each background category in \SRfar and \SRnear, including all uncertainties in the \acp{TF}.
In both regions, muons from heavy-flavour decays dominate the background estimate.

\begin{table}[tp]
\centering
\caption{
Predicted background event yields and expected signal yields for three representative benchmark models in the two signal regions.
The background uncertainties include the statistical and systematic uncertainties in the \acfp{TF}, and statistical uncertainties reflecting the finite event yields in the regions to which the \acp{TF} are applied.
An additional uncertainty is included for the heavy-flavour background in \SRfar, derived from an observed discrepancy between predicted and observed event yields in the $\mDV < \SI{20}{\GeV}$ validation region (see \cref{sec:tf-validation}).
}
\label{tab:result_yields}
\begin{tabular}{l S[table-format=1.2(3)] S[table-format=1.3(4)]}
\toprule
Process & {\SRfar} & {\SRnear} \\
\midrule
Heavy flavour            &  1.2(1.0)    &  2.3(0.8)  \\
Cosmic rays              &  0.18(0.10)  &  0.23(0.13)  \\
Algorithmic fakes barrel &  0.31(0.15)  &  0.31(0.15)  \\
Algorithmic fakes endcap &  0.11(0.06)  &  0.06(0.04)  \\
\addlinespace
Total background         &  1.8(1.1)    &  2.9(0.8)  \\
\midrule
\multicolumn{3}{l}{Higgsino, $\lambda'_{211}$ decays, $m = \SI{1500}{\GeV}$} \\
\quad $\tau = \SI{1}{\ns}$    & 3.4(0.5) & 0.153(0.027) \\
\quad $\tau = \SI{0.1}{\ns}$  & 4.5(0.7) & 0.66(0.06) \\
\quad $\tau = \SI{0.01}{\ns}$ & 0.55(17) & 0.64(15) \\
\bottomrule
\end{tabular}
\end{table}

\subsection{Validation}
\label{sec:tf-validation}

\begin{figure}[tp]
\centering
\includegraphics[width=0.98\linewidth,trim=0 20pt 0 0]{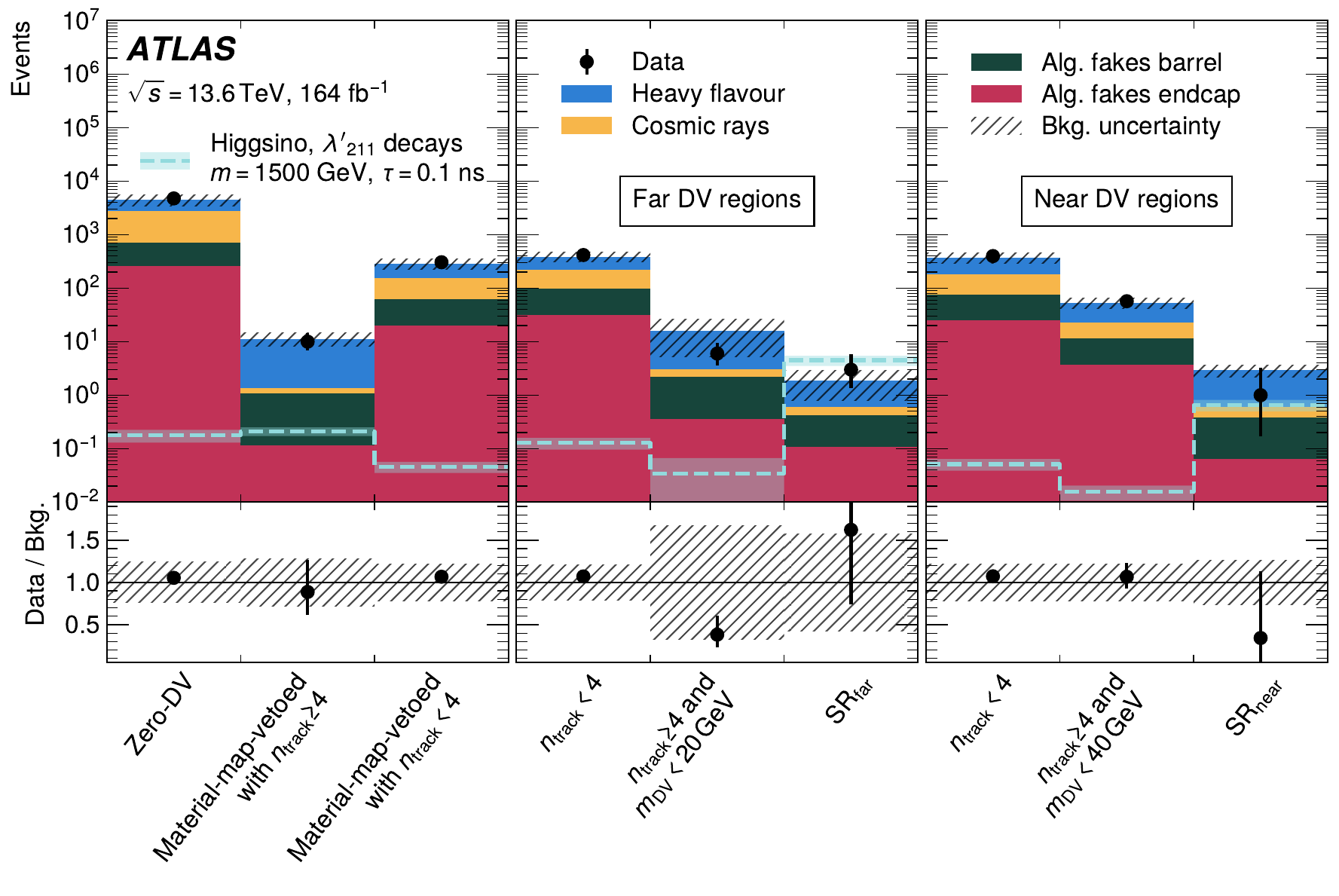}
\caption{%
Overview of the analysis regions with varying \acf{DV} selection criteria.
The predicted background yields are obtained via the \acl{TF} method from the zero-\ac{DV} and material-map-vetoed \ac{DV} selections, with additional muon selection criteria to enhance the purity for each background.
Good agreement is observed in all validation regions, except the far $\ntrack \geq 4, \mDV < \SI{20}{\GeV}$ region.
An additional non-closure uncertainty is applied to the heavy-flavour estimate in that region and in \SRfar.
The hatched band represents the total uncertainty in the prediction.
The predictions are overlaid with one representative higgsino pair production benchmark model, with $m = \SI{1500}{\GeV}$ and $\tau = \SI{0.1}{\ns}$, and \acl{LNV} decays via the $\lambda'_{211}$ coupling.
The uncertainty in the signal prediction includes statistical and systematic uncertainties.
}
\label{fig:region-overview}
\end{figure}

The \acp{TF} are validated in the zero-\ac{DV} and material-map-vetoed \ac{DV} selections, but without applying the restrictive muon selection criteria used for the \ac{TF} derivation in \cref{sec:tf-derivation}.
This provides a direct check that these criteria do not bias the \acp{TF}, and that the systematic uncertainties cover potential residual correlations between the muon-veto requirements and the \ac{DV} selection criteria, and residual background contamination of the \ac{TF} derivation regions.
Good agreement is observed, with all data yields within $1\sigma$ uncertainty of the predictions, demonstrating the robustness of the background estimate.
The agreement between predicted and observed event yields in these validation regions is shown in \cref{fig:region-overview}.

\begin{figure}[tp]
\centering
\subcaptionbox{\label{fig:far-lowNTrk-mu-d0}}{%
\includegraphics[width=0.48\textwidth]{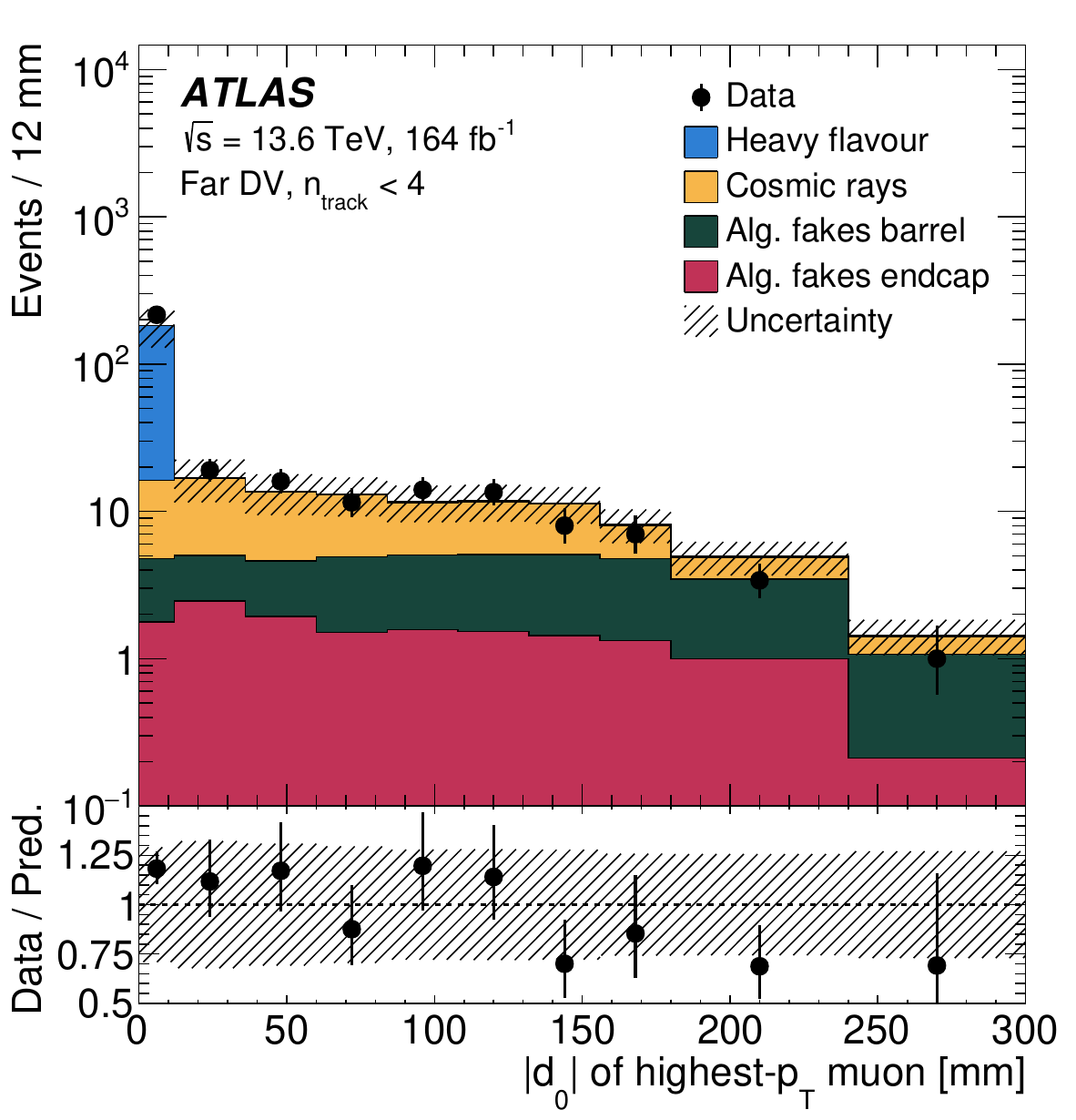}}%
\subcaptionbox{\label{fig:near-lowNTrk-mu-pt}}{%
\includegraphics[width=0.48\textwidth]{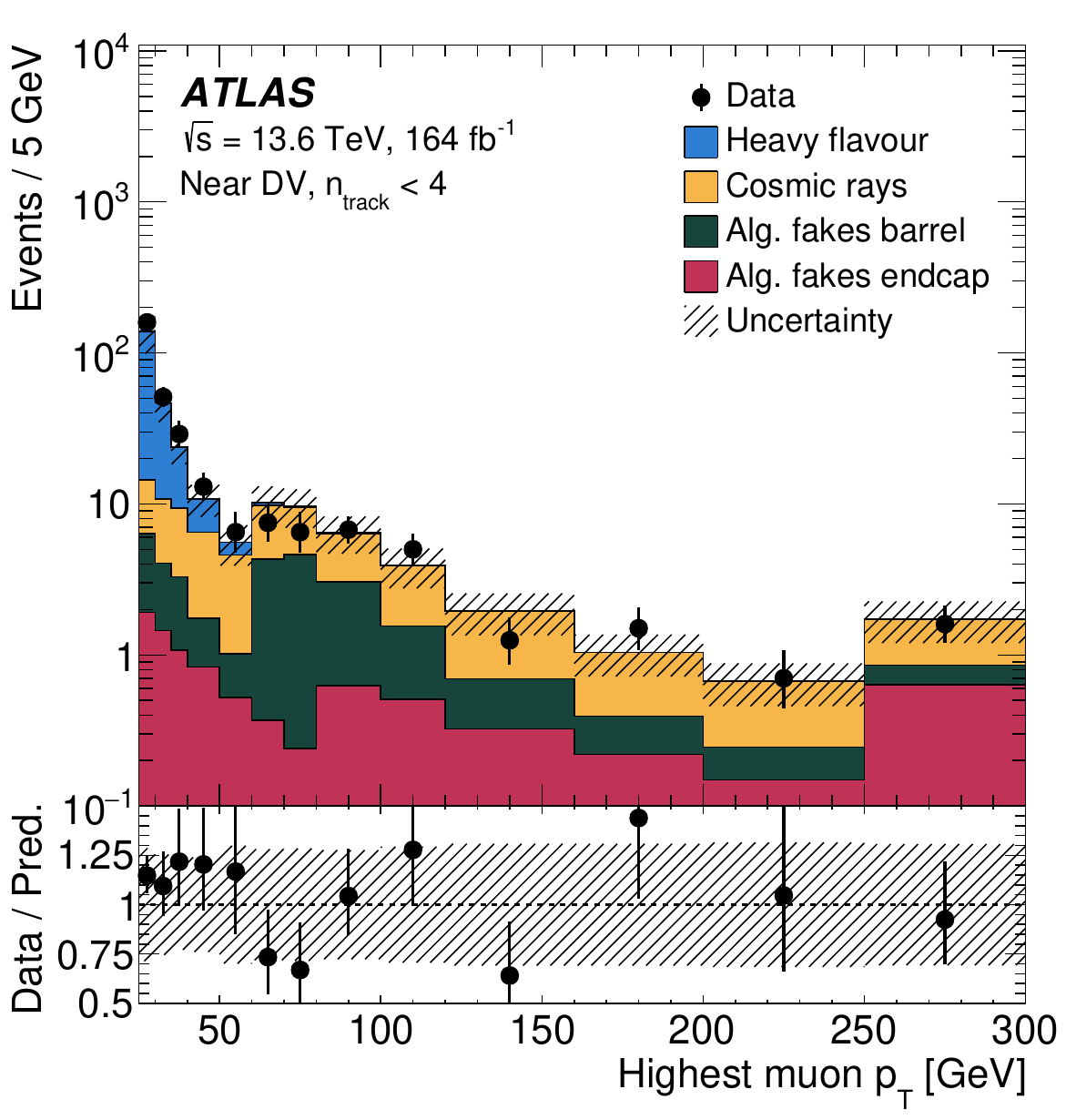}}
\caption{%
Distributions of the predicted background yields and the observed data for events with
\subref{fig:far-lowNTrk-mu-d0}~a far \acf{DV} with a transverse distance from the primary vertex larger than \SI{4}{\mm} and $\ntrack < 4$ as a function of the $|d_0|$ of the highest-\pt muon, and
\subref{fig:near-lowNTrk-mu-pt}~a near \ac{DV} with a transverse distance from the primary vertex between \SI{1}{\mm} and \SI{4}{\mm} and $\ntrack < 4$ as a function of the \pt of the highest-\pt muon.
The hatched band represents the total uncertainty in the prediction.
Overflow events are included in the last bins.
Good agreement is observed in both regions.
}
\label{fig:validation-low-ntrk}
\end{figure}

Further validation is performed by applying the \acp{TF} to the other validation regions outlined in \cref{fig:abcd-schematic},
namely events with \acp{DV} that fail to meet one or multiple signal \ac{DV} criteria:
events with a far \ac{DV} with $\ntrack < 4$,
events with a far \ac{DV} with $\mDV < \SI{20}{\GeV}$,
events with a near \ac{DV} with $\ntrack < 4$, and
events with a near \ac{DV} with $\mDV < \SI{40}{\GeV}$.
Good agreement is observed for all but the far $\mDV < \SI{20}{\GeV}$ region:
while the \ac{TF} method predicts $16 \pm 4$ events, only six are observed.
Although no physical source for a systematic bias was identified, as supported by the good agreement observed in the corresponding near \ac{DV} region with $\mDV < \SI{40}{\GeV}$, a conservative approach is adopted and a non-closure uncertainty is derived for the heavy-flavour background, which dominates this selection.
This uncertainty of approximately 77\% is then applied to the heavy-flavour estimate in this region and propagated to \SRfar.
\Cref{fig:region-overview} also shows the agreement between predicted and observed event yields in these validation regions, with the non-closure uncertainty included where applicable.
\Cref{fig:validation-low-ntrk} shows the distributions of the predicted background yields and the observed data for events with a far \ac{DV} with $\ntrack < 4$ and a near \ac{DV} with $\ntrack < 4$ as a function of the $|d_0|$ of the highest-\pt muon and the \pt of the highest-\pt muon, respectively.

%

\section{Signal Systematic Uncertainties}

Several sources of systematic uncertainty are evaluated for the predicted yields from the signal benchmark models.
These include modelling and instrumental uncertainties, affecting both overall event yields and shapes of distributions used in the statistical analysis described in \cref{sec:results}.
Summing all individual contributions in quadrature, the average total systematic uncertainty in the signal selection efficiency is 24\% and 20\% for \SRnear and \SRfar, respectively.

For all considered signal modelling uncertainties, differences between the overall signal normalisation are removed, such that the quoted uncertainties reflect only acceptance effects.
Uncertainties from missing higher orders in \ac{QCD} in the simulation are evaluated by varying the renormalisation and factorisation scale by factors \num{0.5} and \num{2}.
The variations are done both simultaneously and independently, and the largest deviation from the nominal prediction in each direction is taken as the uncertainty. The average variation over the considered signal samples is 3\% for \SRnear and 2\% for \SRfar.
The maximal variations are 13\% and 9\%, respectively.
Uncertainties owing to the choice of the \ac{PDF} set are evaluated following Ref.~\cite{Butterworth:2015oua} by reweighting the simulated events to the 100 replica sets of the NNPDF \ac{PDF} set and taking the standard deviation of the resulting event yields as the uncertainty.
The average effect on the signal predictions is observed to be 3\% and 5\% in \SRnear and \SRfar, respectively.

The uncertainty in the efficiency for the reconstruction of primary and large-impact-parameter tracks is derived and propagated to the vertexing efficiency, following the methods described in Ref.~\cite{EXOT-2018-57}.
The average effect on the event yields for the different benchmark models is around 2\% and 3\% in \SRnear and \SRfar, respectively.

Uncertainties in the muon reconstruction, isolation and identification efficiencies are computed following the method described in Ref.~\cite{MUON-2018-03}; additional uncertainties in the background veto efficiencies are evaluated.
A dedicated efficiency correction is applied for the identification of displaced muons.
Uncertainties are derived combining two terms: one associated with the reconstruction of the \ac{ID} track, and one associated with the reconstruction in the \ac{MS} segments.
Overall, these correspond to a signal uncertainty between 0.1\% and 2\%.
Uncertainties in the simulation-to-data trigger scale factors are considered for standard prompt muons, and the impact varies between 0.2\% and 8\%;
an additional 3\% uncertainty accounts for the displaced-muon trigger efficiency, computed by comparing the trigger tracking efficiency relative to offline tracks in data and \ac{MC} simulation, as done in Ref.~\cite{TRIG-2022-01}.

The uncertainty associated with the pile-up reweighting procedure~\cite{STDM-2015-05}, averaged over the signal simulations considered, is 3\% and 2\% for \SRnear and \SRfar, respectively.

Furthermore, the expected number of signal events depends on the total integrated luminosity of the data sample.
The combined integrated 2022--2024 luminosity has an uncertainty of 1.9\%, determined using the LUCID--2 detector~\cite{LUCID2}, which provides the primary luminosity measurement, complemented by measurements using the \ac{ID} and the calorimeters.


%

\section{Results}
\label{sec:results}

Three events are observed in \SRfar and one event is observed in \SRnear, fully consistent with the combined background predictions, as summarised in \cref{tab:model_indep_limits}.

\begin{figure}[tp]
\centering
\subcaptionbox{\label{fig:control_plots_SR_far}}{\includegraphics[width=0.48\textwidth]{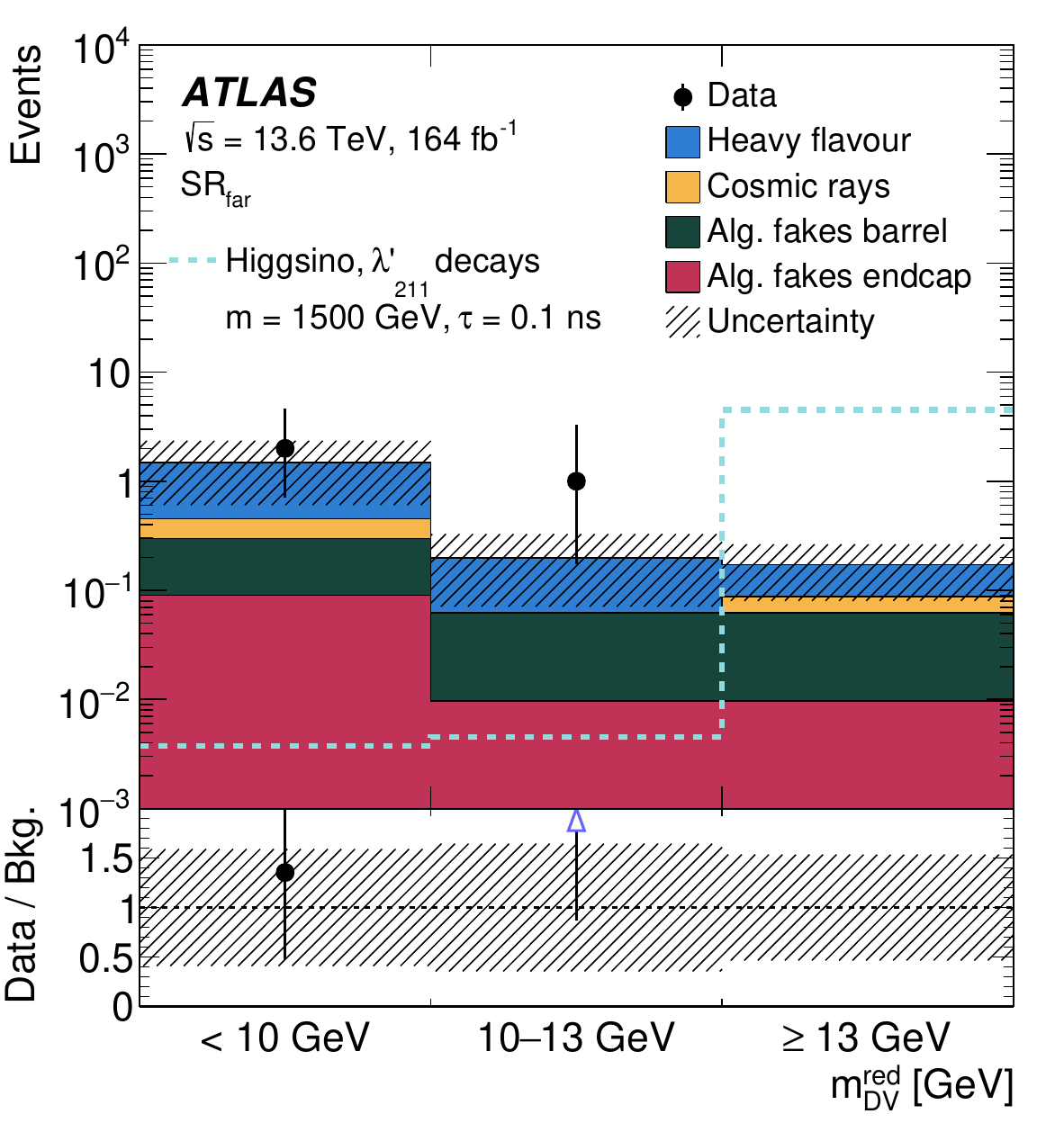}}%
\subcaptionbox{\label{fig:control_plots_SR_near}}{\includegraphics[width=0.48\textwidth]{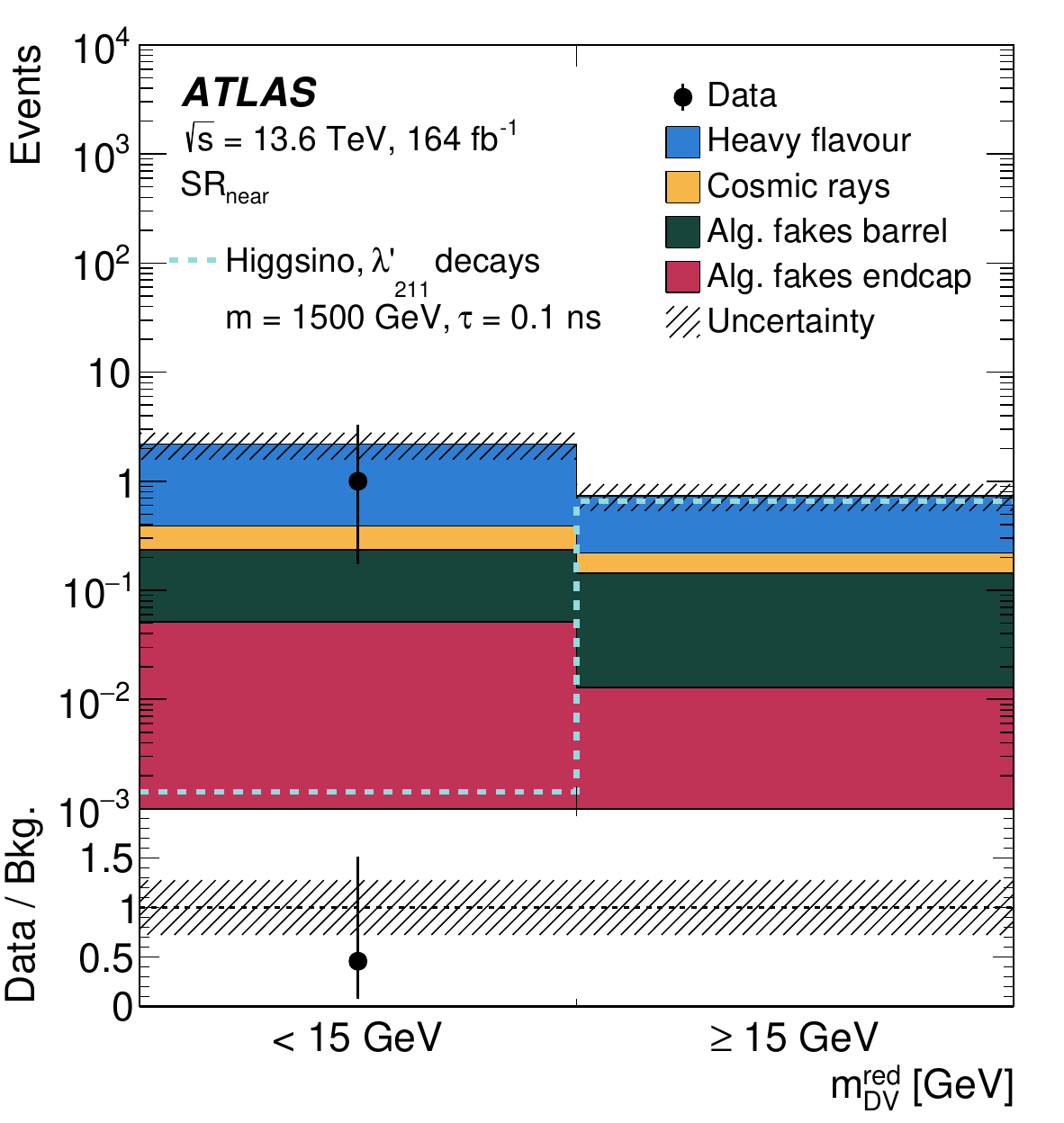}}%
\caption{%
Distributions of the largest reduced invariant mass of any \acl{DV}, \mDVred, in the two signal regions:
\subref{fig:control_plots_SR_far}~\SRfar and \subref{fig:control_plots_SR_near}~\SRnear.
In both signal regions, no events are observed in data in the two highest \mDVred bins.
The background predictions derived using the data-driven \acl{TF} method are shown, along with an example signal benchmark (higgsino pair production with $\lambda'_{211}$ decays, $m = \SI{1500}{\GeV}$ and $\tau = \SI{0.1}{\ns}$) overlaid for illustration.
The lower panels show the ratio of observed data to the total background prediction.
}
\label{fig:control_plots_SR}
\end{figure}

A statistical analysis is performed to derive upper limits on the visible cross-section of final states with \acp{DV} and displaced muons, and to interpret the results in terms of upper limits on the production cross-sections of the benchmark models described in \cref{sec:samples}.
The reduced invariant mass of the \ac{DV}, $\mDVred = \mDV / \Delta R_{\mathrm{max}}$,
where $\Delta R_{\mathrm{max}}$ is the largest angular separation between any pair of \ac{DV}-associated tracks,
is used as the fitting observable.
The average angular separation is typically smaller for signal than for background events, providing additional discrimination over \mDV alone.
In events with more than one signal-like \ac{DV}, only the vertex with the highest \mDVred value is considered.
The best sensitivity to a broad range of signal benchmark models is achieved with three bins in \SRfar, with boundaries at $\mDVred = \SI{10}{\GeV}$ and $\mDVred =$\SI{13}{\GeV}, and two bins in \SRnear, with a boundary at $\mDVred = \SI{15}{\GeV}$.
The \mDVred distributions with this binning in the two signal regions are shown in \cref{fig:control_plots_SR}, overlaid with one example signal benchmark.

The limit extraction is done with the profile likelihood method as implemented in the \AtlasMC{RooFit}, \AtlasMC{HistFactory} and \AtlasMC{TRExFitter} frameworks~\cite{Rolke_2004,RooFit,HistFactory,TRExFitter}.
The signal strength serves as the parameter of interest, while nuisance parameters account for systematic uncertainties in the signal and background predictions and are constrained by Gaussian priors.
The likelihood combines Poisson terms for the event yields in the signal regions and in the regions used to derive the background estimates, to which the \acp{TF} are applied.
The sensitivity of the analysis is primarily limited by the available data statistics, with systematic uncertainties having a subdominant impact.
Consequently, any potential residual mismodelling effects between data and simulation for these exotic signatures have a very limited influence on the final results.
Upper limits at 95\% \ac{CL} are derived using the \CLs method~\cite{Read:2002hq,Junk_1999}, profiling the nuisance parameters to obtain limits on the signal strength.
For the \ac{RPV} interpretations, the likelihood fit incorporates the binned distributions of the discriminating variable, \mDVred, to enhance sensitivity across the benchmark models.

Model-independent 95\% \ac{CL} upper limits on the visible cross-section, $\langle\sigma A\epsilon\rangle$, for final states containing at least one \ac{DV} and at least one displaced muon are obtained from counting experiments performed separately in \SRfar and \SRnear.
In each region, progressively tighter selections are formed by increasing the minimum $\mDVred$ threshold through the exclusion of one bin at a time.
Limits are derived from ensembles of pseudo-experiments using the \CLs method.
\Cref{tab:model_indep_limits} summarises the total background predictions, observed yields, and the resulting $\langle\sigma A\epsilon\rangle^{95}_{\mathrm{obs}}$ for each selection, and the corresponding observed and expected upper limits on the number of signal events.
The most stringent cross-section limits are obtained when only the highest-$\mDVred$ bin of \SRfar\ or \SRnear\ is considered.
The limits on the visible cross-section are of the same order of magnitude as those obtained in the ATLAS Run-2 search for \acp{DV} and displaced muon tracks~\cite{SUSY-2018-33}, but derived in a different fiducial volume.

\begin{table}[tp]
\centering
\caption{%
Model-independent upper limits at 95\% confidence level on the visible cross-section, $\left<\sigma A \epsilon\right>_{\mathrm{obs}}^{95}$, in the two signal regions.
The limits are derived using the \CLs method~\cite{Read:2002hq, Junk_1999} and include the effect of systematic uncertainties as nuisance parameters in the statistical analysis.
The limits are given for three different minimum requirements on the reduced invariant mass of the \acl{DV}, $\mDVred$, in \SRfar and two different minimum requirements in \SRnear.
\emph{Total predicted} lists the total background prediction in each selection, with uncertainties including both statistical and systematic components.
\emph{Observed} lists the number of events observed in data for each selection.
The $\left<\sigma A \epsilon\right>_{\mathrm{obs}}^{95}$ column lists the observed upper limits on the visible cross-section;
the $S_{\mathrm{obs}}^{95}$ and $S_{\mathrm{exp}}^{95}$ columns list the observed and expected upper limits on the number of signal events, respectively.
}
\label{tab:model_indep_limits}
\begin{tabular}{r S[table-format=1.2(3)] S[table-format=1] S[table-format=1.3] S S}
\toprule
& {Total predicted} & {Observed} & {$\left<\sigma A \epsilon\right>_{\mathrm{obs}}^{95}$ [fb]} & {$S_{\mathrm{obs}}^{95}$} & {$S_{\mathrm{exp}}^{95}$} \\
\midrule
\SRfar, $\mDVred$ inclusive            & 1.8(1.1)  & 3 & 0.038 & 6.2 & 4.9 \\
\SRfar, $\mDVred \geq \SI{10}{\GeV}$   & 0.37(0.20) & 1 & 0.023 & 3.8 & 3.1 \\
\SRfar, $\mDVred \geq \SI{13}{\GeV}$   & 0.17(0.09) & 0 & 0.018 & 3.0 & 3.0 \\
\midrule
\SRnear, $\mDVred$ inclusive           & 2.9(0.8)  & 1 & 0.021 & 3.5 & 4.7 \\
\SRnear, $\mDVred \geq \SI{15}{\GeV}$  & 0.73(0.20) & 0 & 0.018 & 3.0 & 3.0 \\
\bottomrule
\end{tabular}
\end{table}

\begin{figure}[tp]
\centering
\subcaptionbox{\label{fig:contour-higgsino-lqd}}{%
\includegraphics[width=0.495\textwidth]{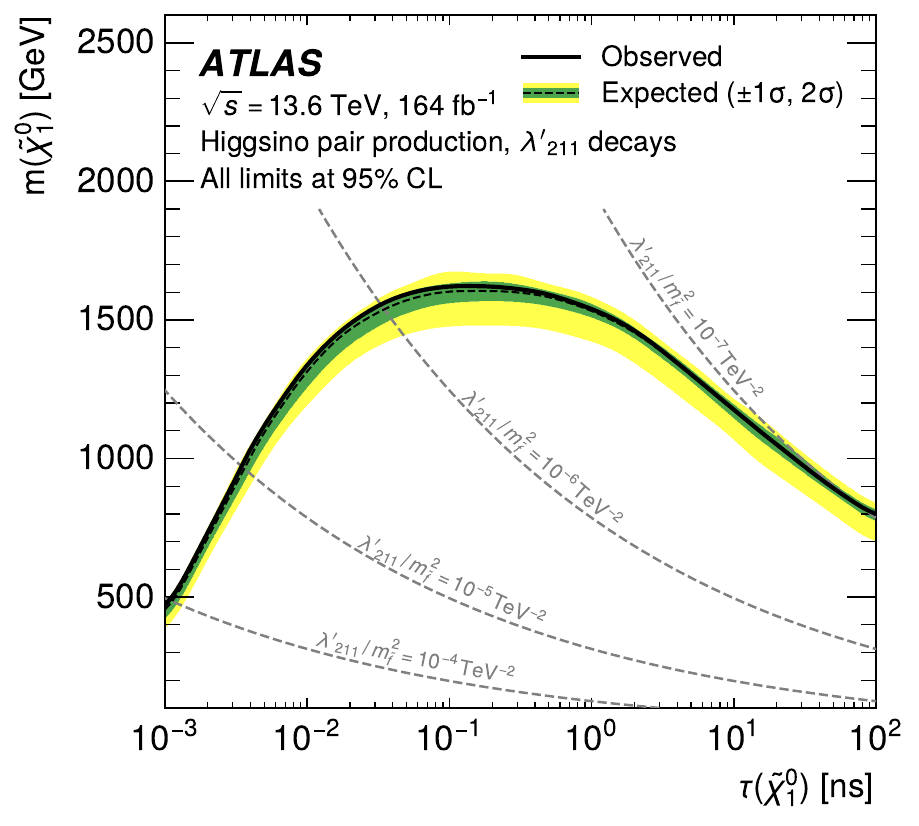}}%
\hfill%
\subcaptionbox{\label{fig:contour-higgsino-udd}}{%
\includegraphics[width=0.495\textwidth]{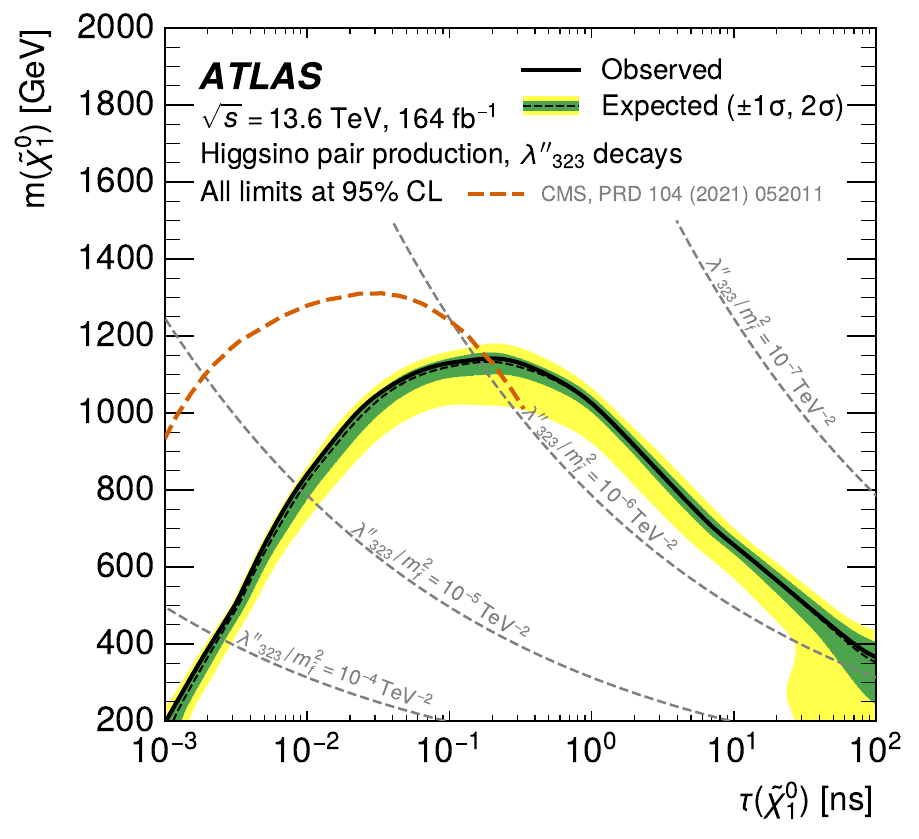}}%
\\
\subcaptionbox{\label{fig:contour-stop-lqd}}{%
\includegraphics[width=0.495\textwidth]{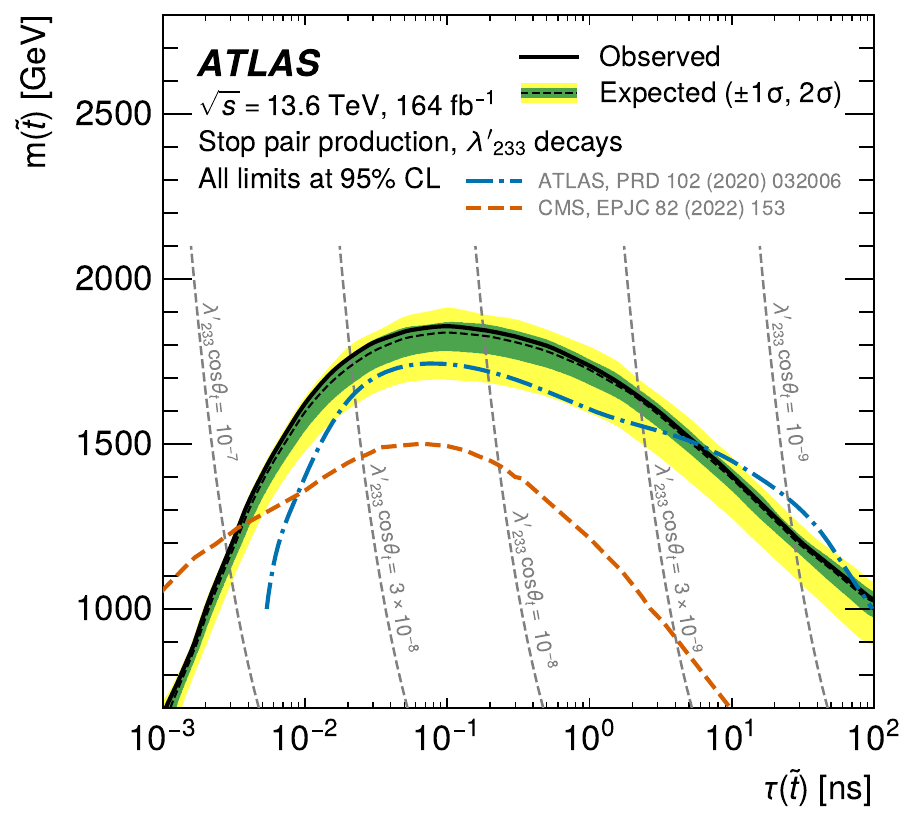}}%
\caption{%
Expected and observed exclusion contours at 95\% confidence level in the plane of the \acl{LLP}'s mass and proper lifetime, for the three considered benchmark models:
higgsino pair production with \subref{fig:contour-higgsino-lqd}~\ac{LNV} $\lambda'_{211}$ and \subref{fig:contour-higgsino-udd}~\ac{BNV} $\lambda''_{323}$ decays,
and \subref{fig:contour-stop-lqd}~$\stoppair$ pair production with \ac{LNV} $\lambda'_{233}$ decays.
A triangular interpolation scheme between simulated and lifetime-reweighted points is used to obtain a smooth contour.
The triangulation is done linearly in mass, and logarithmically in lifetime and signal strength.
Additional smoothing using a Gaussian kernel is applied to the observed limit contour to reduce artifacts from the interpolation.
The contours are overlaid with isocontours of the \acf{RPV} \ac{SUSY} coupling strength.
In the top-squark model, these isocontours correspond to the coupling strength $\lambda'_{233} \cos \theta_t$, where $\theta_t$ is the mixing angle between left- and right-handed top squarks.
In the higgsino models, the values are given for the coupling strengths $\lambda/m_{\tilde{f}}^2$, where $m_{\tilde{f}}$ is the mass of the virtual sfermion mediating the \ac{RPV} decay (with the assumption that $m_{\tilde{f}} \gg m_{\neutralinoOne}$).
The contours in~\subref{fig:contour-higgsino-udd} and~\subref{fig:contour-stop-lqd} are overlaid with results from previous searches for similar models~\cite{CMS-EXO-19-013,SUSY-2018-33,CMS-EXO-18-003}.
}
\label{fig:contours}
\end{figure}

\begin{figure}[tp]
\centering
\subcaptionbox{\label{fig:aux-1dband-LQD}}{%
\includegraphics[width=0.49\linewidth]{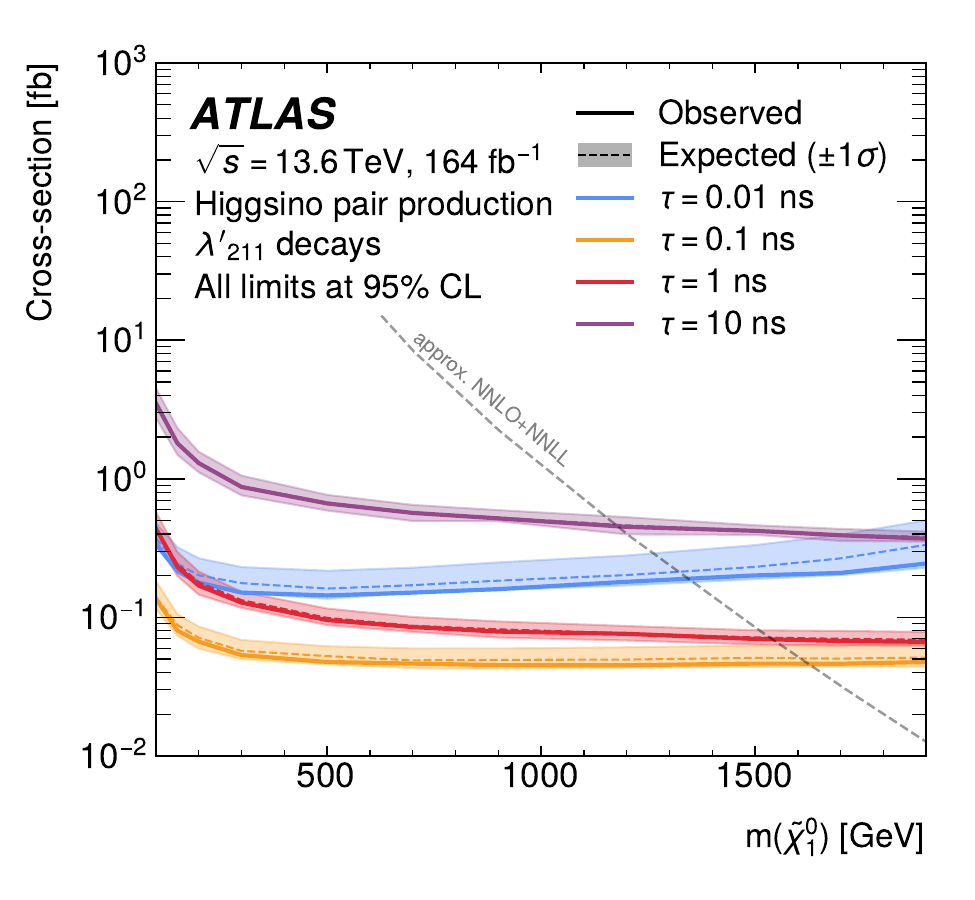}}
\subcaptionbox{\label{fig:aux-1dband-UDDHF}}{%
\includegraphics[width=0.49\linewidth]{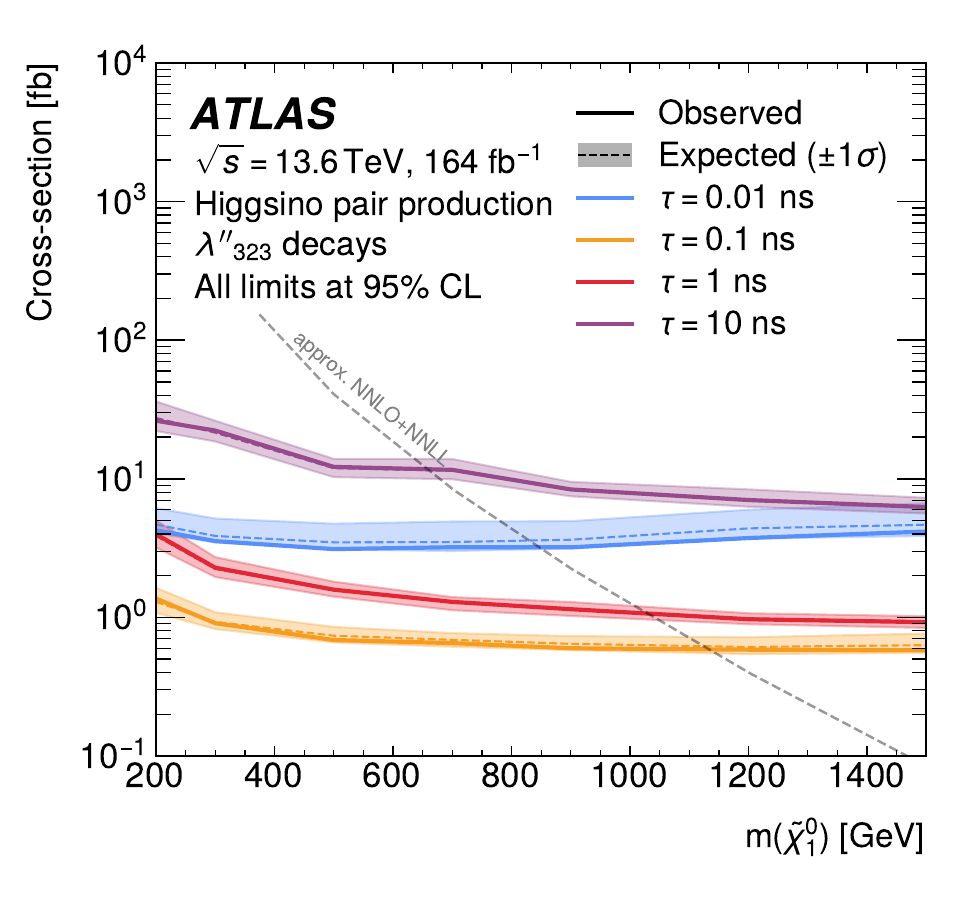}}\\
\subcaptionbox{\label{fig:aux-1dband-stop}}{%
\includegraphics[width=0.49\linewidth]{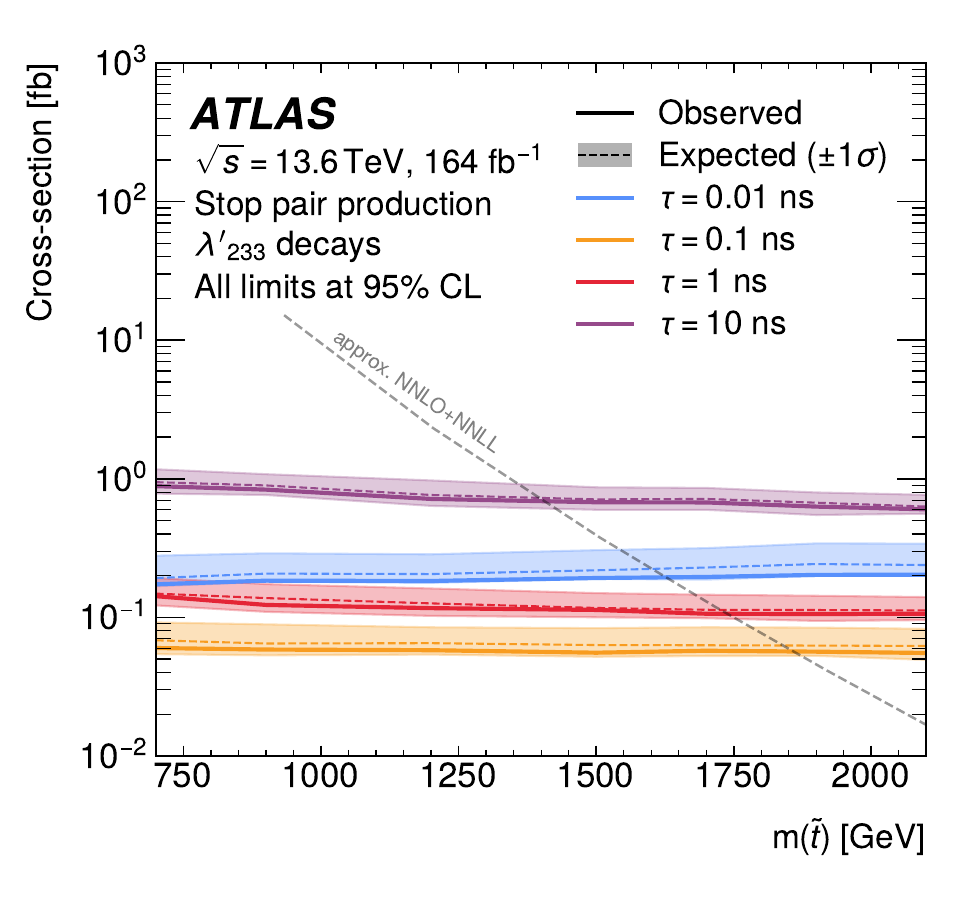}}
\caption{%
Expected and observed 95\% confidence level upper limits on the production cross-section for four representative lifetime values, shown as a function of the \acl{LLP} mass in the higgsino pair production benchmark model with decays via \subref{fig:aux-1dband-LQD}~the $\lambda'_{211}$ coupling and \subref{fig:aux-1dband-UDDHF}~the $\lambda''_{323}$ coupling,
and \subref{fig:aux-1dband-stop}~the top-squark mass in the top-squark model with decays via the $\lambda'_{233}$ coupling.
The expected limits are shown with their one standard deviation uncertainty bands.
The plots are overlaid with the approximate NNLO+NNLL cross-section curves for higgsino and top-squark pair production~\cite{Fiaschi:2020udf,Fuks:2013vua,Fiaschi:2023tkq}.
Good sensitivity is achieved across the entire mass ranges, with the best sensitivity for lifetimes of \SI{0.1}{\ns}.
}
\label{fig:1dbands}
\end{figure}

Model-dependent 95\% \ac{CL} upper limits on the production cross-sections of the benchmark models described in \cref{sec:samples} are obtained from ensembles of pseudo-experiments following a simultaneous fit to \SRfar and \SRnear, using a one-sided profile-likelihood ratio test statistic~\cite{Cowan:2010js} within the \CLs framework.
The signal acceptance times efficiency ($A \times \epsilon$) reaches peak values of approximately 38\% (31\%) for the \ac{LNV} higgsino (top-squark) model for a proper lifetime of $\SI{0.1}{\ns}$, while being about one order of magnitude lower for the \ac{BNV} higgsino model where muons only arise from top-quark decays.
\Cref{fig:contours} shows the expected and observed exclusion contours in the plane of the \ac{LLP} mass and proper lifetime for the three considered benchmark models:
higgsino pair production with (a) \ac{LNV} $\lambda'_{211}$ and (b) \ac{BNV} $\lambda''_{323}$ couplings,
and (c) $\stoppair$ production with \ac{LNV} $\lambda'_{233}$ couplings.
The contours are overlaid with isocontours of the \ac{RPV} \ac{SUSY} coupling strength for each model.
For the top-squark model, these isocontours correspond to the coupling strength $\lambda'_{233} \cos \theta_t$, where $\theta_t$ is the mixing angle between left- and right-handed top squarks, calculated following the expression provided in Ref.~\cite{Kon:1994xe}.
In the higgsino models, the values are given for the coupling strengths $\lambda/m_{\tilde{f}}^2$, where $m_{\tilde{f}}$ is the mass of the virtual sfermion mediating the \ac{RPV} decay, following the parameterisation in Ref.~\cite{Dreiner:1991pe}.
Upper limits on the production cross-sections are also derived for four representative lifetime values, shown as a function of the \ac{LLP} mass in \cref{fig:1dbands}.

Limits on higgsino masses up to \SI{1600}{\GeV} (\SI{1150}{\GeV}) are placed for proper lifetimes of \SI{0.1}{\ns} for $\lambda'_{211}$ ($\lambda''_{323}$) decays,
and higgsino masses between \SI{100}{\GeV} and at least \SI{450}{\GeV} are excluded for proper lifetimes between \SI{1}{\ps} and \SI{100}{\ns} for $\lambda'_{211}$ decays.
Compared with the $\lambda'_{211}$ limits set by the ATLAS Run-1 search for higgsino-like neutralinos~\cite{SUSY-2012-14}, the reach is significantly extended in both mass and proper lifetime while placing fewer assumptions on the production mechanism.
On average, the limits on the production cross-sections for higgsino-like neutralinos with $\lambda'_{211}$ decays are improved by approximately two orders of magnitude.
For $\lambda''_{323}$ decays, the proper lifetime reach is extended up to \SI{100}{\ns}, providing complementary coverage to the CMS Run-2 search for hadronic \ac{DV} pairs~\cite{CMS-EXO-19-013}.
Optimised for vertex pairs within the beam pipe radius of \SI{22}{\mm}, the CMS search sets tighter limits at proper lifetimes below \SI{0.1}{\ns}, whereas this result provides unique sensitivity at larger displacements.

Top-squark masses of up to \SI{1850}{\GeV} are excluded for proper lifetimes of \SI{0.1}{\ns}, and any values between \SI{700}{\GeV} and \SI{1500}{\GeV} are excluded for proper lifetimes between \SI{10}{\ps} and \SI{5}{\ns}.
This improves upon the limits set by the previous ATLAS search for \acp{DV} and displaced muon tracks using Run-2 data~\cite{SUSY-2018-33}, particularly for proper lifetimes below \SI{5}{\ns}, highlighting the enhanced sensitivity provided by the new dedicated muon triggers and improved reconstruction techniques employed.
For longer lifetimes, the Run-2 search retains more stringent limits due to the usage of missing transverse momentum triggers, which were not considered in this Letter.

%
\FloatBarrier

\section{Conclusion}

A search for \acp{LLP} decaying into a \ac{DV} and a muon is presented, using \SI{164}{\ifb} of $\sqrt{s} = \SI{13.6}{\TeV}$ $pp$ collision data collected by the ATLAS detector at the LHC between 2022 and 2024.
The search targets events with a \ac{DV} with at least four associated tracks and an invariant mass above \SI{20}{\GeV}, along with at least one muon with a transverse impact parameter $|d_0| > \SI{2}{\mm}$.
A data-driven \ac{TF} method is employed to estimate the background contributions from heavy-flavour decays, cosmic-ray muons, and algorithmic fakes.
Two signal regions are defined based on the transverse distance between the \ac{DV} and the primary interaction vertex; in both regions, the background is dominated by muons from heavy-flavour decays.
No significant excess above the expected background is observed.
The results are interpreted in terms of upper limits on the visible cross-section for final states with at least one \ac{DV} and at least one displaced muon, and as upper limits on the production cross-sections of benchmark models involving \acp{LLP} from \ac{RPV} \ac{SUSY}.
The sensitivity is significantly enhanced over previous searches: cross-section limits for higgsino-like neutralinos with \ac{LNV} $\lambda'_{211}$ decays are improved by approximately two orders of magnitude relative to Run-1 measurements, and the reach for \ac{BNV} $\lambda''_{323}$ decays is extended to proper lifetimes of \SI{100}{\ns}.
For top squarks, the dedicated Run-3 muon triggers and refined reconstruction techniques lead to better sensitivity to $\lambda'_{233} \cos \theta_t$ for proper lifetimes below \SI{5}{\ns} than achieved in prior ATLAS results.


%
\section*{Acknowledgements}

%
%

%
%

We thank CERN for the very successful operation of the LHC and its injectors, as well as the support staff at
CERN and at our institutions worldwide without whom ATLAS could not be operated efficiently.

The crucial computing support from all WLCG partners is acknowledged gratefully, in particular from CERN, the ATLAS Tier-1 facilities at TRIUMF/SFU (Canada), NDGF (Denmark, Norway, Sweden), CC-IN2P3 (France), KIT/GridKA (Germany), INFN-CNAF (Italy), NL-T1 (Netherlands), PIC (Spain), RAL (UK) and BNL (USA), the Tier-2 facilities worldwide and large non-WLCG resource providers. Major contributors of computing resources are listed in Ref.~\cite{ATL-SOFT-PUB-2026-001}.

We gratefully acknowledge the support of ANPCyT, Argentina; YerPhI, Armenia; ARC, Australia; BMWFW and FWF, Austria; ANAS, Azerbaijan; CNPq and FAPESP, Brazil; NSERC, NRC and CFI, Canada; CERN; ANID, Chile; CAS, MOST and NSFC, China; Minciencias, Colombia; MEYS CR, Czech Republic; DNRF and DNSRC, Denmark; IN2P3-CNRS and CEA-DRF/IRFU, France; SRNSFG, Georgia; BMFTR, HGF and MPG, Germany; GSRI, Greece; RGC and Hong Kong SAR, China; ICHEP and Academy of Sciences and Humanities, Israel; INFN, Italy; MEXT and JSPS, Japan; CNRST, Morocco; NWO, Netherlands; RCN, Norway; MNiSW, Poland; FCT, Portugal; MNE/IFA, Romania; MSTDI, Serbia; MSSR, Slovakia; ARIS and MVZI, Slovenia; DSI/NRF, South Africa; MICIU/AEI, Spain; SRC and Wallenberg Foundation, Sweden; SERI, SNSF and Cantons of Bern and Geneva, Switzerland; NSTC, Taipei; TENMAK, T\"urkiye; STFC/UKRI, United Kingdom; DOE and NSF, United States of America.

Individual groups and members have received support from BCKDF, CANARIE, CRC and DRAC, Canada; CERN-CZ, FORTE and PRIMUS, Czech Republic; COST, ERC, ERDF, Horizon 2020 and Marie Sk{\l}odowska-Curie Actions, European Union; Investissements d'Avenir Labex, Investissements d'Avenir Idex and ANR, France; DFG and AvH Foundation, Germany; Herakleitos, Thales and Aristeia programmes co-financed by EU-ESF and the Greek NSRF, Greece; BSF-NSF and MINERVA, Israel; NCN and NAWA, Poland; La Caixa Banking Foundation, CERCA and AGAUR programs from Generalitat de Catalunya and PROMETEO and GenT Programmes Generalitat Valenciana, Spain; G\"{o}ran Gustafssons Stiftelse, Sweden; The Royal Society and Leverhulme Trust, United Kingdom; Eric and Wendy Schmidt Fund for Strategic Innovation, United States of America.

In addition, individual members wish to acknowledge support from Chile: Agencia Nacional de Investigaci\'on y Desarrollo (ANID FONDECYT reg. 1230987, FONDECYT 1230812, FONDECYT 1240864, Fondecyt 3240661, Fondecyt Regular 1240721); China: Chinese Ministry of Science and Technology (MOST-2023YFA1605700, MOST-2023YFA1609300), National Natural Science Foundation of China (NSFC 12275265, NSFC-W2543005); Czech Republic: Czech Science Foundation (GACR - 24-11373S), Ministry of Education Youth and Sports (ERC-CZ-LL2327, FORTE CZ.02.01.01/00/22\_008/0004632), PRIMUS Research Programme (PRIMUS/21/SCI/017); EU: H2020 European Research Council (ERC - 101002463); European Union: European Research Council (BARD No. 101116429, ERC - 948254, ERC 101089007), European Regional Development Fund (HE COFUND GA No.101081355, ERDF), Marie Sklodowska-Curie Actions (GAP-101168829); France: Agence Nationale de la Recherche (ANR-21-CE31-0013, ANR-22-EDIR-0002, ANR-24-CE31-0504-01); Germany: Deutsche Forschungsgemeinschaft (DFG - 469666862); China: Research Grants Council (GRF); Italy: Ministero dell'Università e della Ricerca (NextGenEU 153D23001490006 M4C2.1.1, NextGenEU I53D23000820006 M4C2.1.1, NextGenEU I53D23001490006 M4C2.1.1, SOE2024\_0000023); Japan: Japan Society for the Promotion of Science (JSPS KAKENHI  JP25H0063, JSPS KAKENHI JP22H01227, JSPS KAKENHI JP22H04944, JSPS KAKENHI JP22KK0227, JSPS KAKENHI JP24K23939, JSPS KAKENHI JP24KK0251, JSPS KAKENHI JP25H00650, JSPS KAKENHI JP25H01291, JSPS KAKENHI JP25K01011, JSPS KAKENHI JP25K01023); Norway: Research Council of Norway (RCN-314472); Poland: Polish National Science Centre (NCN 2021/42/E/ST2/00350, NCN OPUS 2023/51/B/ST2/02507, NCN OPUS nr 2022/47/B/ST2/03059, NCN UMO-2019/34/E/ST2/00393, UMO-2022/47/O/ST2/00148, UMO-2023/49/B/ST2/04085, UMO-2023/51/B/ST2/00920, UMO-2024/53/N/ST2/00869); Spain: Agència de Gestió d'Ajuts Universitaris i de Recerca. (AGAUR - 2023 BP 00141), Ministry of Science and Innovation (RYC2019-028510-I, RYC2020-030254-I, RYC2021-031273-I, RYC2022-038164-I), Ministerio de Ciencia, Innovación y Universidades/Agencia Estatal de Investigaci\'on (EU NextGenerationEU (PRTR-C17.I1), PID2022-142604OB-C22); Sweden: Carl Trygger Foundation (Carl Trygger Foundation CTS 22:2312), Swedish Research Council (Swedish Research Council 2023-04654, VR 2021-03651, VR 2022-03845, VR 2022-04683, VR 2023-03403, VR 2024-05451, VR 2025-05940), Knut and Alice Wallenberg Foundation (KAW 2023.0366); Switzerland: Swiss National Science Foundation (SNSF - PCEFP2\_194658); United Kingdom: The Binks Trust, Royal Society (NIF-R1-231091); United States of America: U.S. Department of Energy (ECA DE-AC02-76SF00515), John Templeton Foundation (John Templeton Foundation 63206), Neubauer Family Foundation.

%
%


%
\printbibliography
\clearpage
\input{atlas_authlist}
%

%

\end{document}

%% file: atlas_authlist.tex
 
\begin{flushleft}
\hypersetup{urlcolor=black}
{\Large The ATLAS Collaboration}

\bigskip

\AtlasOrcid[0000-0002-6665-4934]{G.~Aad}$^\textrm{\scriptsize 102}$,
\AtlasOrcid[0000-0001-7616-1554]{E.~Aakvaag}$^\textrm{\scriptsize 17}$,
\AtlasOrcid[0000-0002-5888-2734]{B.~Abbott}$^\textrm{\scriptsize 121}$,
\AtlasOrcid[0000-0002-0287-5869]{S.~Abdelhameed}$^\textrm{\scriptsize 83b}$,
\AtlasOrcid[0000-0002-1002-1652]{K.~Abeling}$^\textrm{\scriptsize 54}$,
\AtlasOrcid[0000-0001-5763-2760]{N.J.~Abicht}$^\textrm{\scriptsize 48}$,
\AtlasOrcid[0000-0002-8496-9294]{S.H.~Abidi}$^\textrm{\scriptsize 30}$,
\AtlasOrcid[0009-0003-6578-220X]{M.~Aboelela}$^\textrm{\scriptsize 44}$,
\AtlasOrcid[0000-0002-9987-2292]{A.~Aboulhorma}$^\textrm{\scriptsize 36e}$,
\AtlasOrcid[0000-0001-5329-6640]{H.~Abramowicz}$^\textrm{\scriptsize 154}$
% ...
\bigskip
\\

% ...
$^{17}$Department for Physics and Technology, University of Bergen, Bergen; Norway.\\
% ...
$^{30}$Physics Department, Brookhaven National Laboratory, Upton NY; United States of America.\\
% ...
$^{36}$$^{(a)}$Facult\'e des Sciences Ain Chock, Universit\'e Hassan II de Casablanca;$^{(b)}$Facult\'{e} des Sciences, Universit\'{e} Ibn-Tofail, K\'{e}nitra;$^{(c)}$Facult\'e des Sciences Semlalia, Universit\'e Cadi Ayyad, LPHEA-Marrakech;$^{(d)}$LPMR, Facult\'e des Sciences, Universit\'e Mohamed Premier, Oujda;$^{(e)}$Facult\'e des sciences, Universit\'e Mohammed V, Rabat;$^{(f)}$Institute of Applied Physics, Mohammed VI Polytechnic University, Ben Guerir; Morocco.\\
% ...
$^{44}$Physics Department, Southern Methodist University, Dallas TX; United States of America.\\
% ...
$^{48}$Fakult\"{a}t Physik, Technische Universit{\"a}t Dortmund, Dortmund; Germany.\\
% ...
$^{54}$II. Physikalisches Institut, Georg-August-Universit\"{a}t G\"ottingen, G\"ottingen; Germany.\\
% ...
$^{83}$$^{(a)}$Khalifa University of Science and Technology, Abu Dhabi;$^{(b)}$New York University Abu Dhabi, Abu Dhabi;$^{(c)}$United Arab Emirates University, Al Ain;$^{(d)}$University of Sharjah, Sharjah; United Arab Emirates.\\
% ...
$^{102}$CPPM, Aix-Marseille Universit\'e, CNRS/IN2P3, Marseille; France.\\
% ...
$^{121}$Homer L. Dodge Department of Physics and Astronomy, University of Oklahoma, Norman OK; United States of America.\\
% ...
$^{154}$Raymond and Beverly Sackler School of Physics and Astronomy, Tel Aviv University, Tel Aviv; Israel.\\
% ...
$^{a}$ Also at Affiliated with an institute formerly covered by a cooperation agreement with CERN.\\
$^{b}$ Also at An-Najah National University, Nablus; Palestine.\\
$^{c}$ Also at Borough of Manhattan Community College, City University of New York, New York NY; United States of America.\\
$^{d}$ Also at Center for Interdisciplinary Research and Innovation (CIRI-AUTH), Thessaloniki; Greece.\\
$^{e}$ Also at Centre of Physics of the Universities of Minho and Porto (CF-UM-UP); Portugal.\\
$^{f}$ Also at CERN, Geneva; Switzerland.\\
% ...

\end{flushleft}
